\documentclass[epjST]{svjour}
\usepackage{latexsym}
\usepackage{graphics}
\usepackage{amssymb}
\newcommand{\beq}{\begin{equation}}
\newcommand{\eeq}{\end{equation}}

\newcommand{\bea}{\begin{eqnarray}}
\newcommand{\eea}{\end{eqnarray}}
\begin{document}
\title{Complexity Methods Applied to Turbulence in Plasma Astrophysics}
\author{Loukas Vlahos \and Heinz Isliker}
%
%
\institute{Department of Physics, Aristotle University, 54124 Thessaloniki, Greece}
\date{Received: date / Revised version: date}
%
\abstract{
In this review many of the well known tools for the analysis of Complex systems are used in order to study the global coupling of the turbulent convection zone with the solar atmosphere where the magnetic energy is dissipated explosively. Several well documented observations are not easy to interpret with the use of Magnetohydrodynamic (MHD)  and/or Kinetic numerical codes. Such observations are:  (1) The size distribution of the Active Regions (AR) on the solar surface, (2) The fractal and multi fractal characteristics of the observed magnetograms, (3) The Self-Organised characteristics of the explosive magnetic energy release and (4) the very efficient acceleration of particles during the flaring periods in the solar corona. We review briefly the work published  the last twenty five years on the above issues and propose solutions by using methods borrowed from the analysis of complex systems. The scenario which emerged is as follows: (a) The fully developed turbulence in the convection zone generates and transports magnetic flux tubes to the solar surface. Using probabilistic percolation models we were able to reproduce the size distribution and the fractal properties of the emerged and randomly moving magnetic flux tubes. (b) Using a Non Linear Force Free (NLFF) magnetic extrapolation numerical code we can explore how the emerged magnetic flux tubes interact nonlinearly and form thin and Unstable Current Sheets (UCS) inside the coronal part of the AR. (c) The fragmentation of the UCS and the redistribution of the magnetic field locally, when the local current exceeds a Critical threshold, is a key process which drives avalanches and  forms coherent structures. This local reorganization of the magnetic field, enhances the energy dissipation and influences the global evolution of the complex magnetic topology. Using a Cellular Automaton and follow the simple rules of  Self Organized Criticality (SOC), we were able to reproduce the statistical characteristics of the observed time series of the explosive events, (d) finally, when the AR reaches the {\bf turbulently reconnecting state} (in the language of the SOC theory this is called {\bf SOC state}) it is densely populated by UCS which can act as local scatterers replacing the ({\bf magnetic clouds}) in the Fermi scenario and enhance dramatically the heating and acceleration of charged particles.}

%
\maketitle
\section{Introduction}
\label{intro}
A rapidly growing area of scientific inquiry is the exploration of the dynamics of complex systems. A defining characteristic of complex systems is their tendency to self-organize globally as a result of many local interactions. In other words, organization occurs without any central organizing structure or entity. Such self-organization has been observed in systems at scales from neurons to ecosystems.
A complex adaptive system has the following characteristics: it persists in spite of changes in the diverse individual components of which it is comprised; the interactions between those components are responsible for the persistence of the system; and the system itself engages in adaptation or learning (\cite{Holland}, pp.4). To say that a system is complex is to say that it moves between order and disorder without becoming fixed in either state. To say that such a system adapts is to say that it responds to information by changing.

There are systems that persist in spite of the continual changes of individual components, maintaining coherence and adapting in response to a phenomenal amount of information throughout the lifetime of the organism in which they function (\cite{Holland}, pp. 2-3).

The process by which a complex system achieves adaption results in self-organization by the system, that is, {\bf agents} acting locally, unaware of the extent of the larger system of which they are a part, generate larger patterns which result in the organization of the system as a whole. Not every system is a complex adaptive system; certain conditions must be met in order for a system to self-organize. First of all, the system must include a large number of agents. In addition, the agents must interact in a nonlinear fashion (see \cite{Kauffman,Kelso}).

Magnetohydrodynamic (MHD)  turbulence is a very common phenomenon in many astrophysical and laboratory plasma systems. Although the driver which is responsible for the fluctuations may be different, the end product is always the same and depends on the energy carried by the possibly unstable magnetic fluctuations. When the level of the unstable fluctuations ($\delta B$)   is small in comparison with the the ambient magnetic filed $B$ ($\delta B/B<<1$) the turbulence is called {\bf weak turbulence}  and  systems driven to large amplitude magnetic fluctuations  ($\delta B/ B \geq  1$) are called  {\bf strong turbulence (or fully developed turbulence).}  The main difference between {\bf weak} and {\bf strong}  turbulence is the formation of coherent structures (Eddies and/or Unstable Current Sheets) which play a crucial role in the dissipation of the magnetic energy in {\bf strong turbulence}.  In weak turbulence the unstable waves loose energy by  interacting  quasi-linearly with the plasma particles. In this article, we will be addressing phenomena related mostly with {\bf strong turbulence,} which will be called simply  {\bf turbulence.}

Many astrophysical and laboratory systems are in the turbulent state. Such well known systems are: e.g. the convection zone of the Sun and stars,  the solar wind, Earth's magnetotail,  Earth's Bow shock, the supernova remnants,   the accretion discs and the relativistic jets around compact objects, super clusters of galaxies, edge turbulence in laboratory plasmas, etc. The driver which brings all the above systems into a turbulent state is different but the systems they share several properties which will be discussed in this review.

The formation of coherent structures (Eddies and Unstable Current Sheets) inside turbulent systems, has been investigated with many different numerical and analytical tools. In this article we will concentrate on numerical tools which are used extensively in the analysis of complex systems. The most popular numerical tool used to analyse complex systems is the Cellular Automaton (CA) \cite{Chop}. The set up for the CA depends strongly on the qualitative analysis of the physical system under study, which defines the {\bf rules}  of the CA. The succes of the outcome of the CA is tested with the direct comparison with the data and parallel  MHD simulations. So complex systems, like turbulent plasmas, can be explored by CA models (on the global astrophysical scales) and MHD or kinetic simulations for the local scales. The MHD and Kinetic simulations can serve as tools for testing the {\bf rules} of the CA.

In this review we will focus on phenomena related to the strong coupling of the turbulent convection zone with the atmosphere of the Sun, and more precisely  the regions on the solar surface which host strong magnetic field patterns and are called Active Regions (AR). The convection zone is the outer-most layer of the solar interior. It extends from a depth of 200,000 km up to the visible surface of the Sun.  The coherent structures (eddies) inside the convection zone generate and transport towards the surface numerous thin loops (also called threads, fibrils and etc.) The thin loops are  analogous to the {\bf agents} mentioned earlier. The thin loops generated in the turbulent convection zone are numerous and they interact nonlinearly to form coherent magnetic patterns in the solar photosphere (analogous to the coherent structures in ecosystems) and drive the heating and acceleration of particles in the complex magnetic topology above the photosphere by forming current sheets at different scales.

In this review we address three important questions wich are closely related to the nonlinear magnetic coupling of the turbulent convection zone with the AR.

\begin{itemize}
	\item How the thin magnetic loops form the observed coherent magnetic patterns in the solar photosphere?
	\item How the driven complex magnetic topology above the photosphere is forming numerous thin current sheets (some of them unstable) and drive the observed Self-Organized  explosive phenomena?
	\item How the the Self-Organised turbulent current sheets heat and accelerate the plasma inside the AR?
\end{itemize}

In section \ref{s:2} we apply the percolation theory to simulate the formation of the photospheric part of the AR,
in section \ref{sec:3} we discuss briefly  the role of Self Organized Criticality (SOC) in solar physics and the interpretation of the explosive patterns, in section \ref{sec:4}
we discuss the heating and acceleration of particles in turbulently reconnecting astrophysical systems and in section \ref{sec:5} we discuss our  main conclusions.

\section{Percolation theory in astrophysics: Formation of solar active regions \label{s:2}}
\subsection{Statistical analysis of the solar magnetograms}
\begin{figure}[ht]
  \centering
  \resizebox{0.45\columnwidth}{!}{ \includegraphics{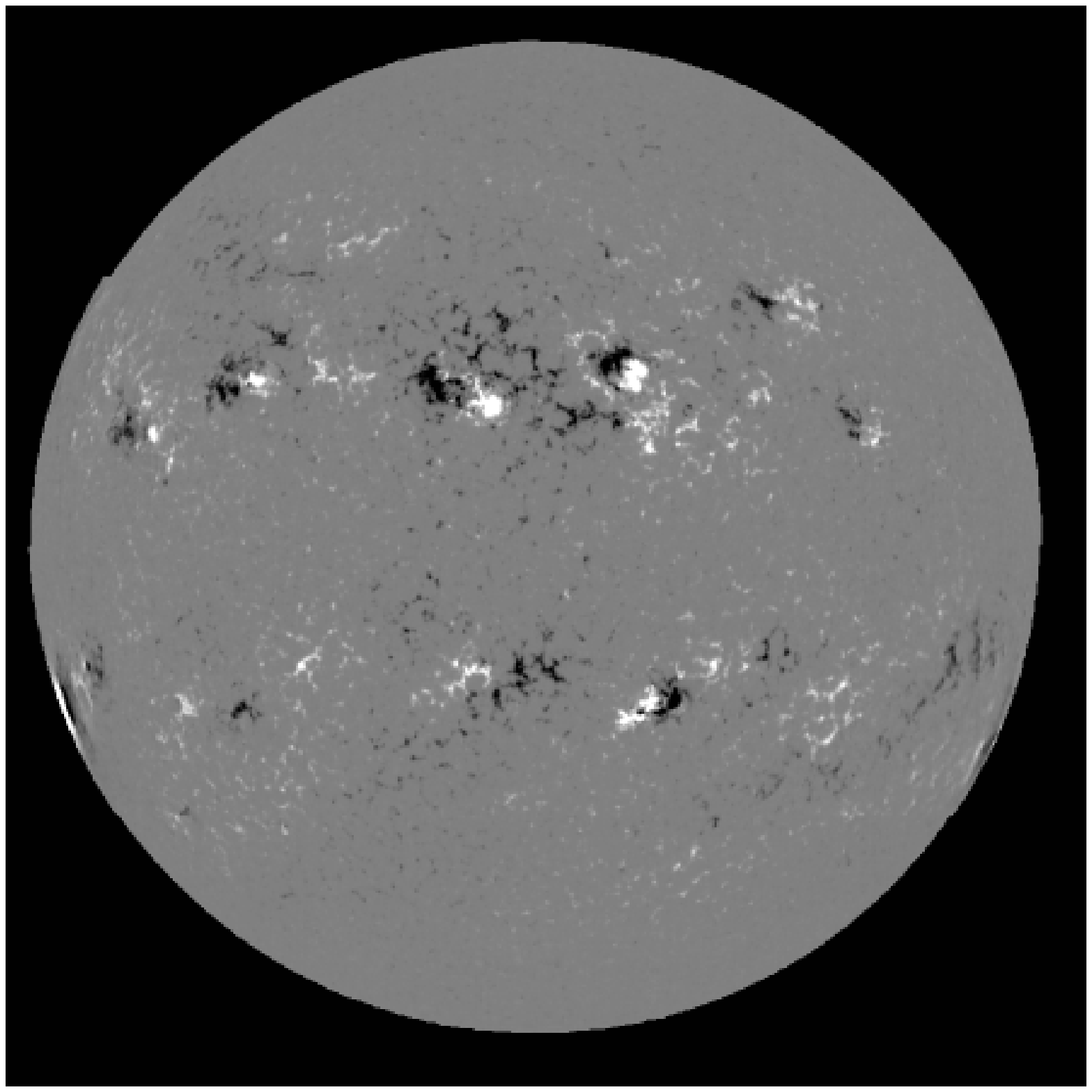}}
  \resizebox{0.45\columnwidth}{!}{ \includegraphics{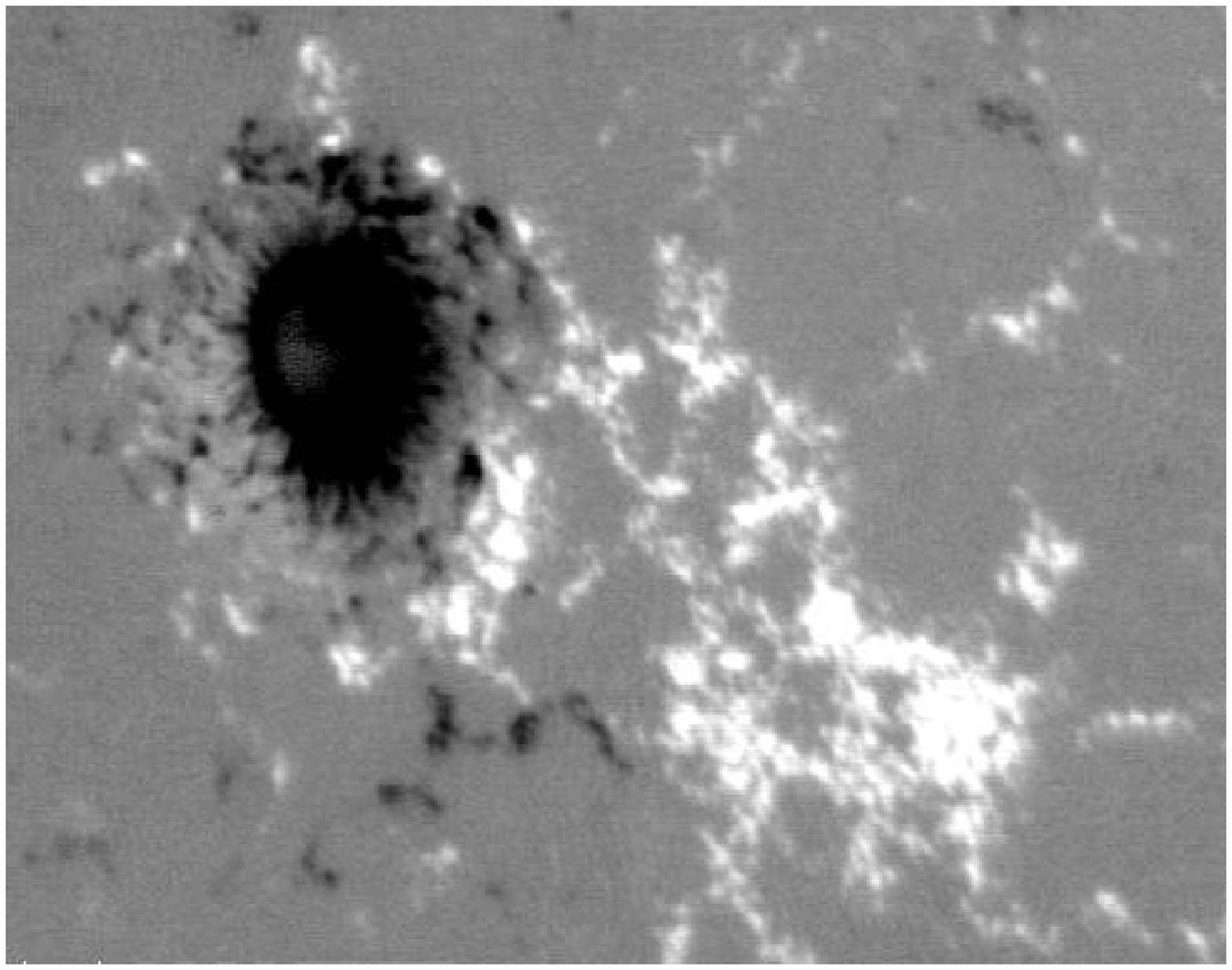}}
\caption{(a) A full disk magnetogram. Several AR are present in the solar surface (b) Zoom on the details of one AR. \label{mag}}
\end{figure}

In Fig. \ref{mag}  we present a typical magnetogram. Magnetograms reproduce the strength and location of the magnetic fields on the Sun. In a magnetogram, grey areas indicate that there is no magnetic field, while black and white areas indicate regions where there is a strong magnetic field.

Numerous observational studies have investigated the statistical
properties of active regions, using full-disc magnetograms.
\begin{figure}[ht]
  \centering
  \resizebox{0.40\columnwidth}{!}{ \includegraphics{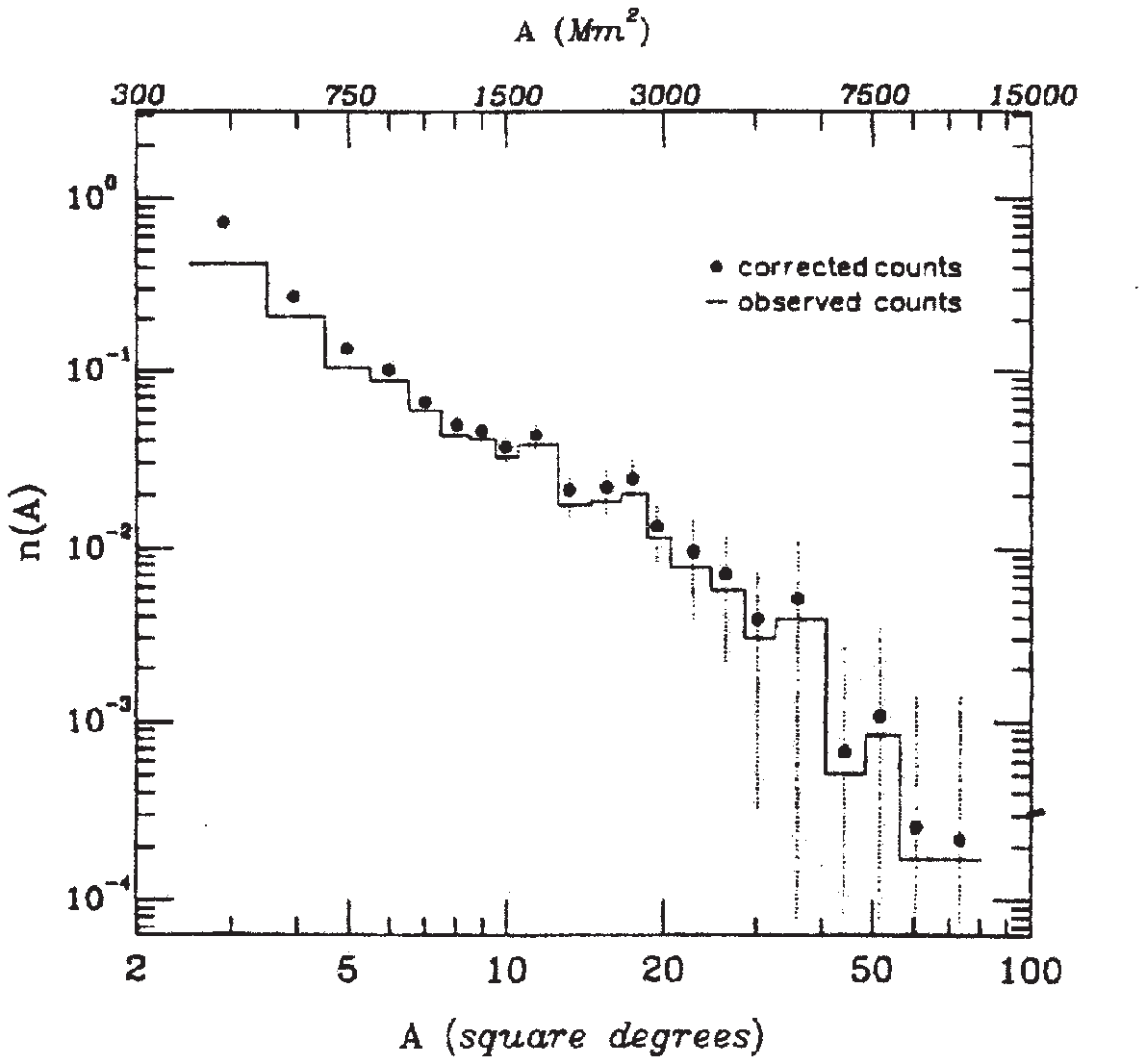}}
  \resizebox{0.50\columnwidth}{!}{ \includegraphics{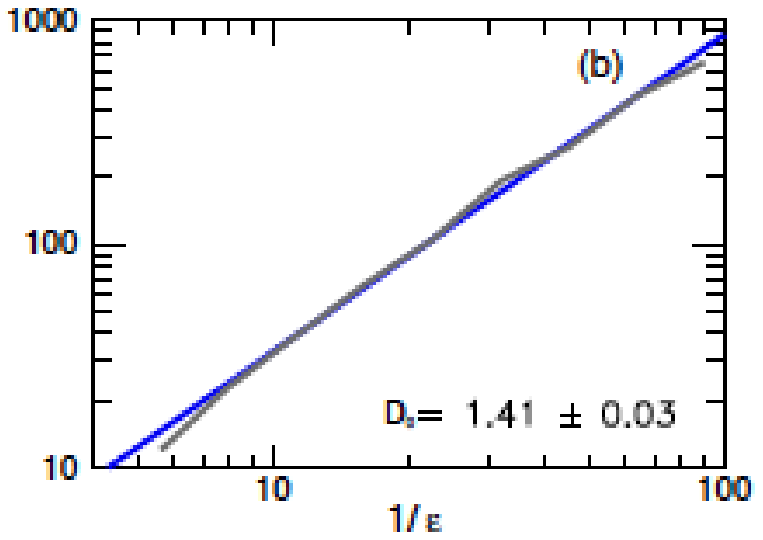}}

  \caption{(a) The size distribution of young active regions \cite{Hzw93} (b) The fractal Dimension $D_0$ from a typical active region magnetogram, where $\epsilon$ is a dimensional variable related with the size of the boxes covering the AR.\cite{Georgoulis2012} \label{s3}}
\end{figure}
These studies have examined among other parameters the Probability Distribution Function (PDF) of the size
distribution of  ARs, and their fractal dimension: {\bf
The size distribution function} of the newly formed ARs
exhibits a well defined power law with index $\approx -1.94$ (see
Figure  \ref{s3}a), and ARs cover only a small
fraction of the solar surface (around $\sim 8\%$) \cite{Hzw93}.
  The fractal dimension and the multifractal structure function spectrum of the
active regions has been studied using high-resolution magnetograms \cite{Balk93,Meun99,Georgoulis2012}. These
authors found, using not always the same method, a fractal
dimension $D_0$ in the range $1.2<D_0<1.7$ (see Fig. \ref{s3}b).
The fractal dimensions for the solar magnetic fields are typically calculated using the box-counting technique. The values of the fractal dimension also depends on whether the structures themselves or just their boundaries are box counted. The analysis has been pursued even further using the concepts of multi-fractality \cite{McAteer,Abramenko,Lawrence}. It is well known that an AR includes multiple types of structures such as different classes of sunspots, plages, emerging flux sub-regions, etc. The physics behind the formation and evolution of each of these structures is not believed to be the same, so the impact and the final outcome of the convection zone turbulence in each of these structures should not be the same. Numerous other tools have been used to uncover aspects of the complex behavior ``mapped" by the convection zone onto the photospheric boundary,  e.g. generalized correlation dimension, structure formations, wavelet power spectrum (see \cite{geo05}).

\subsection{Formation of Active Regions through  the turbulent diffusion of magnetic flux tubes}

The evolution of a single flux tube from the overshoot layer,
where it is generated till it reaches the photosphere, is a very
difficult and challenging problem.  Studies of the dynamics of a
1-D (slender) flux tube are useful since they permit calculation
of the evolution as the flux-tube propagates through the
convection zone. The 2-D and 3-D characteristics of the rise of
magnetic flux tubes allow us to understand the role of the twist
in the properties of a rising flux-tube \cite{Amari,kliem,Archontis,Archontis04,Archontis09,Fan09,aulanie1,Gal3,Galsgaard05}.  A review of the
main results on the evolution of 1-D, 2-D and 3-D flux tubes was
given by Moreno-Insertis \cite{Moreno97,Fanr}. It is well documented from numerical
simulations that  twisted flux tubes can rise almost without change
through the convection zone, if the azimuthal magnetic field is
strong. Two very important questions should be addressed: (1) How
are flux-tubes formed from a large scale magnetic field, and (2)
which is the origin of the twist in the flux tubes.  Hughes et al. \cite{Hug97}
address this problem, their main conclusion is that the non-linear
evolution of the magnetic Rayleigh-Taylor instability is
responsible for the formation of magnetic flux tubes. The origin
of the twist is a much more complex problem. There are several
possibilities: (1) the flux-tubes are formed with a large twist in
the overshoot layer, (2) the twist is built-up  during the
propagation inside the convection zone. It is still an open
question if the required twist can be there from ``birth" or is
added later. We are far from understanding the details both of the
formation and the origin of the twist. Large scale 3-D simulations
of the formation and evolution of magnetic flux tubes will solve
many of the open problems that still exist.

We move from the single flux-tube scenario to the idea  that
magnetic flux-tubes of all sizes and twists are generated in the
overshoot layer and propagate towards the photosphere. The idea to
study the formation of active regions as the outcome of the
statistical evolution of $N$ randomly moving flux-tubes was
proposed by Bogdan \cite{Bog84} and developed further in
\cite{Bog85,BogLer85}. Similar studies on the
statistical mechanics of a gas of vortices, embedded in a
two-dimensional inviscid fluid, were performed by Fr\"{o}hlich $\&$  Ruelle \cite{FrRu82}.
The evolution of a collection of $N$ flux-tubes has the potential to address the
statistical properties of the observed data but it has remained in the definition stage since there are several free parameters which cannot be controled by observations.

We just noted that also several models have been developed using the anomalous diffusion
of magnetic flux in the solar photosphere in order to explain the
fractal geometry of the active regions
\cite{Sch92,Law91,Lsr93,Mze93}.

\subsection{Models based on Percolation theory}

Almost forty years ago, Seiden and Wentzel \cite{Sei96,Wen92} developed a percolation model to simulate the formation and evolution of active regions. In this model, the evolution of active
regions is followed by reducing all the complicated solar MHD and
turbulence to three dimensionless parameters. This percolation
model explained the observed size distribution of active regions
and their fractal characteristics \cite{Meun99}.
Vlahos et al. \cite{vla02} developed further the percolation model for the
emergence and evolution of magnetic flux on the solar surface
using a 2-D cellular automaton (CA), following techniques
developed initially by Seiden and Wentzel \cite{Sei96}. The dynamics of this
automaton is probabilistic and is based on the competition between
two ``fighting" tendencies: \textbf{stimulated} or
\textbf{spontaneous} emergence of new magnetic flux, and the
disappearance of flux due to \textbf{diffusion} (i.e.\ dilution
below observable limits),
 together with random \textbf{motion} of the
flux tubes on the solar surface. The basic new element they added to
the initial model \cite{Sei96} was that they kept track of the
\textbf{energy release} through flux cancellation (reconnection)
if flux tubes of opposite polarities collide. They concentrated
their analysis only on the newly formed active regions, since the
old active regions undergo more complicated behavior.

Following Vlahos et al. \cite{vla02} the main physical properties of active regions, as derived from
the observations of the evolving active regions, can be summarized
in simple CA rules: A 2-D quadratic grid with $200 \times 1000$
cells (grid sites) is constructed, in which each cell has four
nearest neighbors. The grid is assumed to represent a large
fraction of the solar surface. Initially, a small, randomly chosen
percentage ($1\%$) of the cells is magnetized (loaded with flux)
in the form of positively (+1) and negatively (-1) magnetized
pairs (dipoles), the rest of the grid points are set to zero.
Positive and negative cells evolve independently after their
formation, but their percentage remains statistically equal. The
dynamical evolution of the model is controlled by the following
probabilities:

\vspace{0.2cm}

\noindent  {\bf P}: The probability that a magnetized  cell is
stimulating the appearance of new flux at one of its nearest
neighbors. Each magnetized cell can stimulate its neighbors only
the first time step of its life. This procedure simulates the
stimulated emergence of flux which occurs due to the observed
tendency of magnetic flux to emerge in regions of the solar
surface in which magnetic flux had previously emerged.

\noindent {$\bf D_m$}: The flux of each magnetized cell has a
probability $D_m$ to move to a random neighboring cell, simulating
motions forced by the turbulent dynamics of the underlying
convection zone. If the moving flux meets oppositely polarized
flux in a neighboring cell, the fluxes cancel (through
reconnection), giving rise to a ``flare". If equal polarities meet
in a motion event, the fluxes simply add up.
\begin{figure}[ht]
\centering
\resizebox{0.70\columnwidth}{!}{\%\includegraphics{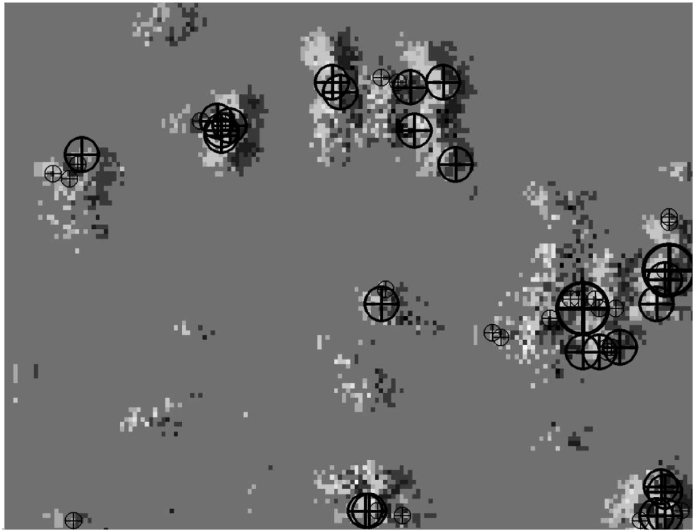}}
\caption{A small portion of the modeled grid is presented. The
dark areas represent negative and the white ones positive magnetic
flux. The explosions (``flares'') appear randomly at the interface
of regions of oppositely polarized magnetic flux. The circles
represent  the positions of the ``flares", with their radius being
proportional to the logarithm of the released energy \cite{vla02}  \label{fig2}}
\end{figure}

\noindent  {$\bf D_d$}: The probability that a magnetized cell is
turned into non-magnetized in one time-step if it is next to a
non-magnetized cell.
This rule simulates two effects, the direct submersion of magnetic
flux and the disappearing of flux below observational limits due
to dilution caused by diffusion into the empty neighborhood.

\noindent {$\mathcal{E}$}: The probability that a non-magnetized
cell is turned into magnetized spontaneously, independently of its
neighbors, simulating the observed spontaneous emergence of new
flux. Every newly appearing flux tube is accompanied by an
oppositely polarized mate, taking into account the fact that flux
appears always in the form of dipoles.

\vspace{0.2cm}

A detailed discussion of the connection between the parameters
$P,D_d,\mathcal{E}$ and the physical mechanisms acting in the
evolution of active regions was established
\cite{Sei96,Wen92}. Vlahos et al. \cite{vla02} performed a series of numerical experiments using
the above model. The parameters used for the results reported here
are $P=0.185, D_d=0.005,$ $D_m=0.05$ and $\mathcal{E}=10^{-6}$.
{\bf They are chosen such that, when following the evolution of our
model and recording the percentage of magnetized cells, we find
that it takes around 1000 time steps before the percentage of
active cells is stabilized to a value which is close to the
observed one (around $8\%$)}. In Fig. \ref{fig2}, we present a
small portion of the grid. Dark areas correspond to negative and
bright ones to positive polarity. The spatial location of the
``flares" is marked with circles, with the size of the circles
proportional to the logarithm of the locally released energy.

The size distribution of the simulated active regions is estimated
and is approximated by a power law fit of the form $N(s)\sim
s^{-k} $, with $k=1.93\pm 0.08$. Finally, we estimate the fractal
dimension $D_0$ of the set of magnetized cells with the box
counting algorithm \cite{Fal90}, finding $D_0=1.42 \pm 0.12$. Both these results are well in accordance with the observations.

The cancellation of magnetic flux due to collisions of oppositely
polarized magnetic flux tubes leads to the release of energy,
whose amount we assume to be proportional to the difference in the
square of the magnetic flux before and after the event (as an approximation of the differences in magnetic energy).
\begin{figure}[ht]
\centering
\resizebox{0.45\columnwidth}{!}{\%\includegraphics{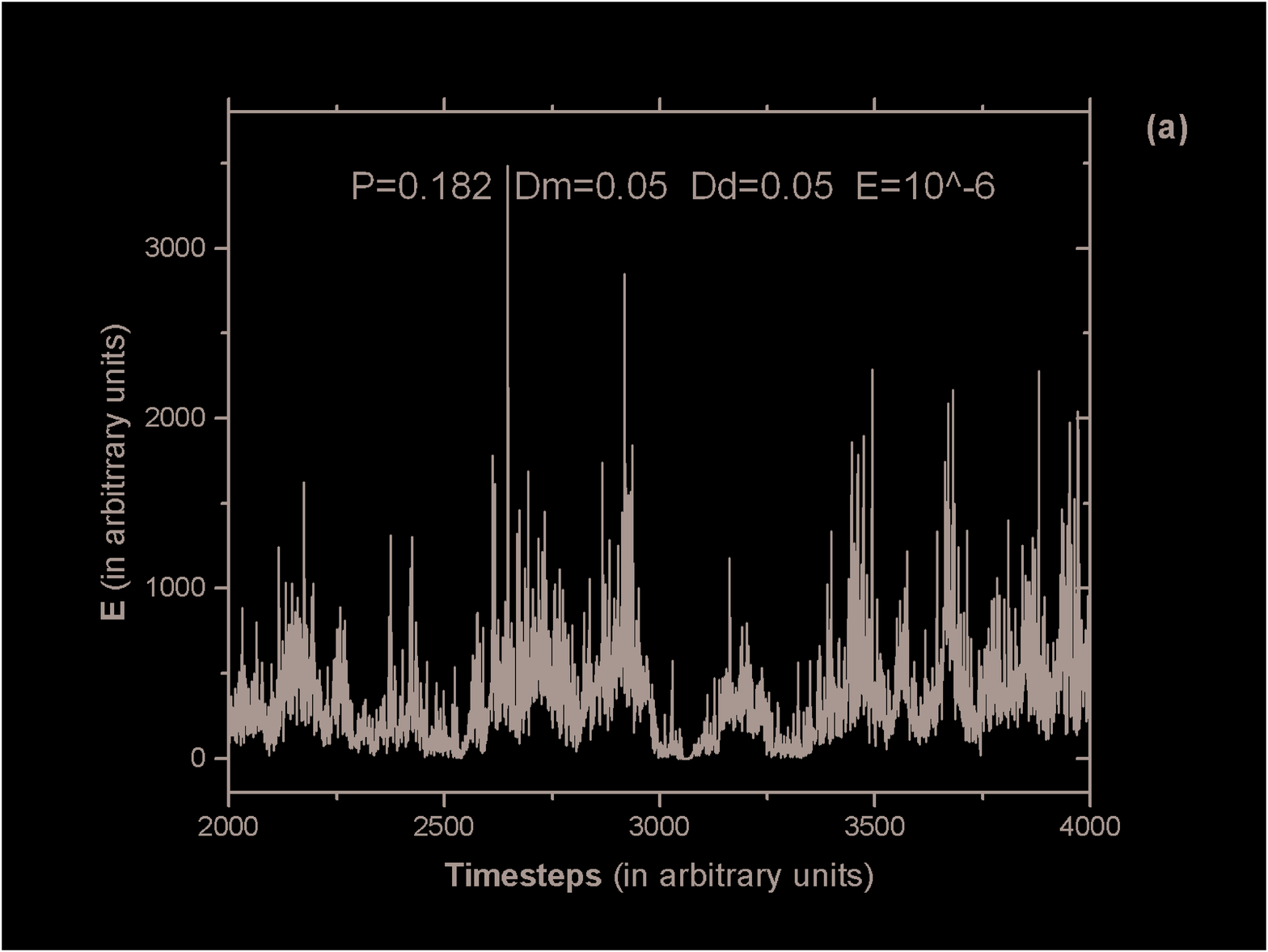}}
\resizebox{0.45\columnwidth}{!}{\%\includegraphics{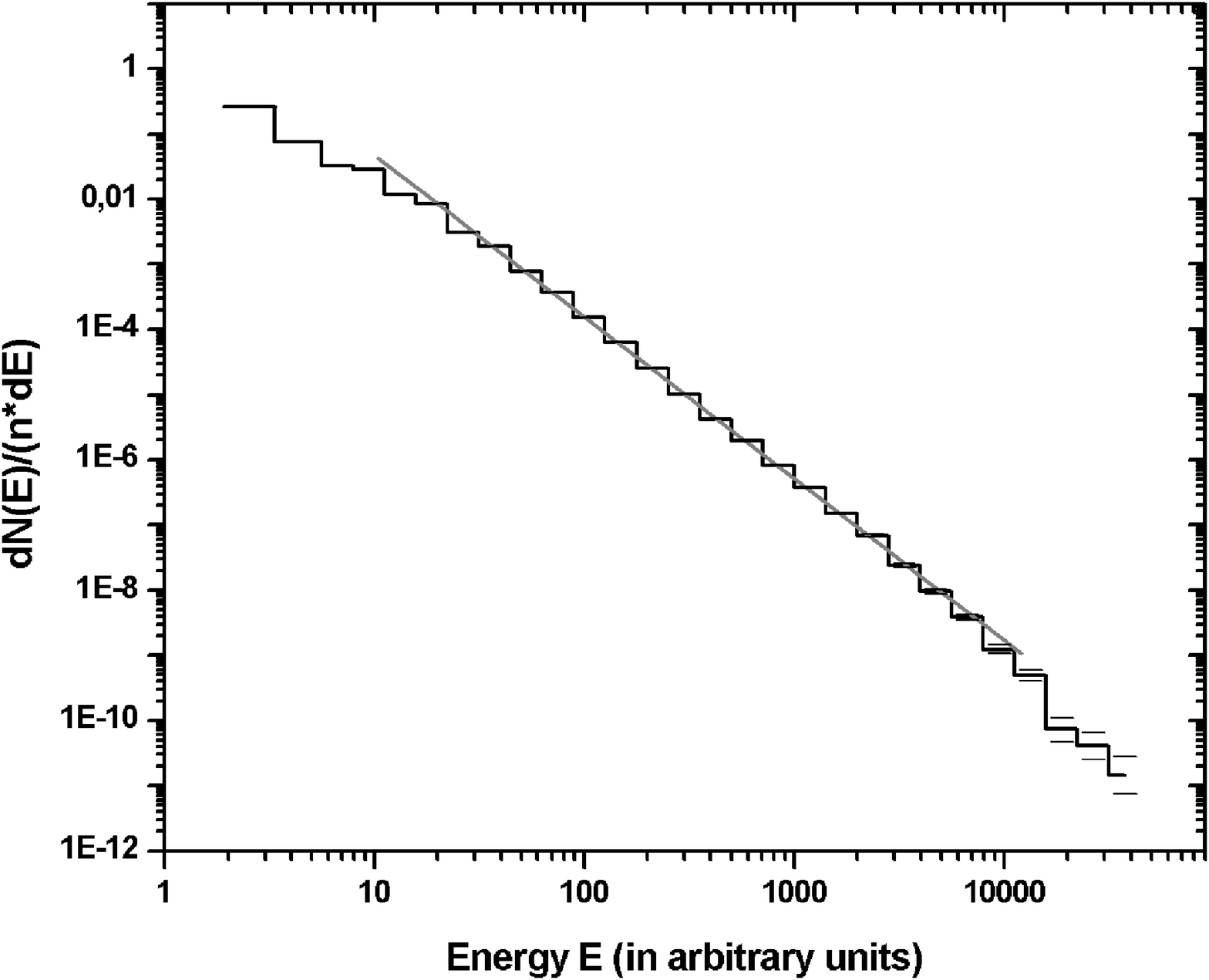}}
\caption{(a) The energy released in the cancellation of magnetic
flux as a function of time, using the parameters $P=1.185,
D_d=0.005, D_m=0.05, \mathcal{E}=10^{-6}$. (b) The energy
distribution of the recorded ``flares". The power-law index is
$2.5\pm 0.13$. \cite{vla02} \label{can}
}
\end{figure}

In Fig. ~\ref{can}a, we plot the released energy $E(t)$ as a
function of time. Fig.~\ref{can}b shows the energy distribution of
the recorded ``explosions": It follows a power law, $f(E)\sim E^{-a}$,
with $a=2.5 \pm 0.13,$ for energies $E>20$. For energies $E<20$
finite resolution effects must be expected to bias the
distribution, so we do not draw conclusions for the small
energies. It is important to note also that the power law in the
distribution of energy is extended over three decades.

A variation of the parameters $P,D_d,D_m$ does not alter the power
law behaviours and the fractality, they seem to be
\textbf{generic} properties of the model. The exact values of the
parameters, $k,a,D_0$ depend though on the free parameters but
remain inside the observed limits even for a large variation of
$P,D_d,D_m$. The results are also independent of $\mathcal{E}$ as
long as it remains small enough.

It has been pointed out  \cite{Georgoulis2012} that the global statistical properties of the ARs discussed above cannot serve as a tool for predicting flaring or non flaring activity on the Sun, since the multiscale AR properties underline the self-organization driven by turbulence, means intermittency in the systems response, hence lack of predictability. In other words, it is not possible to use multiscale measures that exemplify intermittency, or stochasticity, for prediction purposes.

\section{Explosive phenomena in solar active regions: The role of the turbulent driver and Self-Organized Criticality}
\label{sec:3}
\subsection{Preliminaries and the statistical properties of explosive phenomena}
\begin{figure}[ht]
\centering
\resizebox{0.45\columnwidth}{!}{\%
\includegraphics{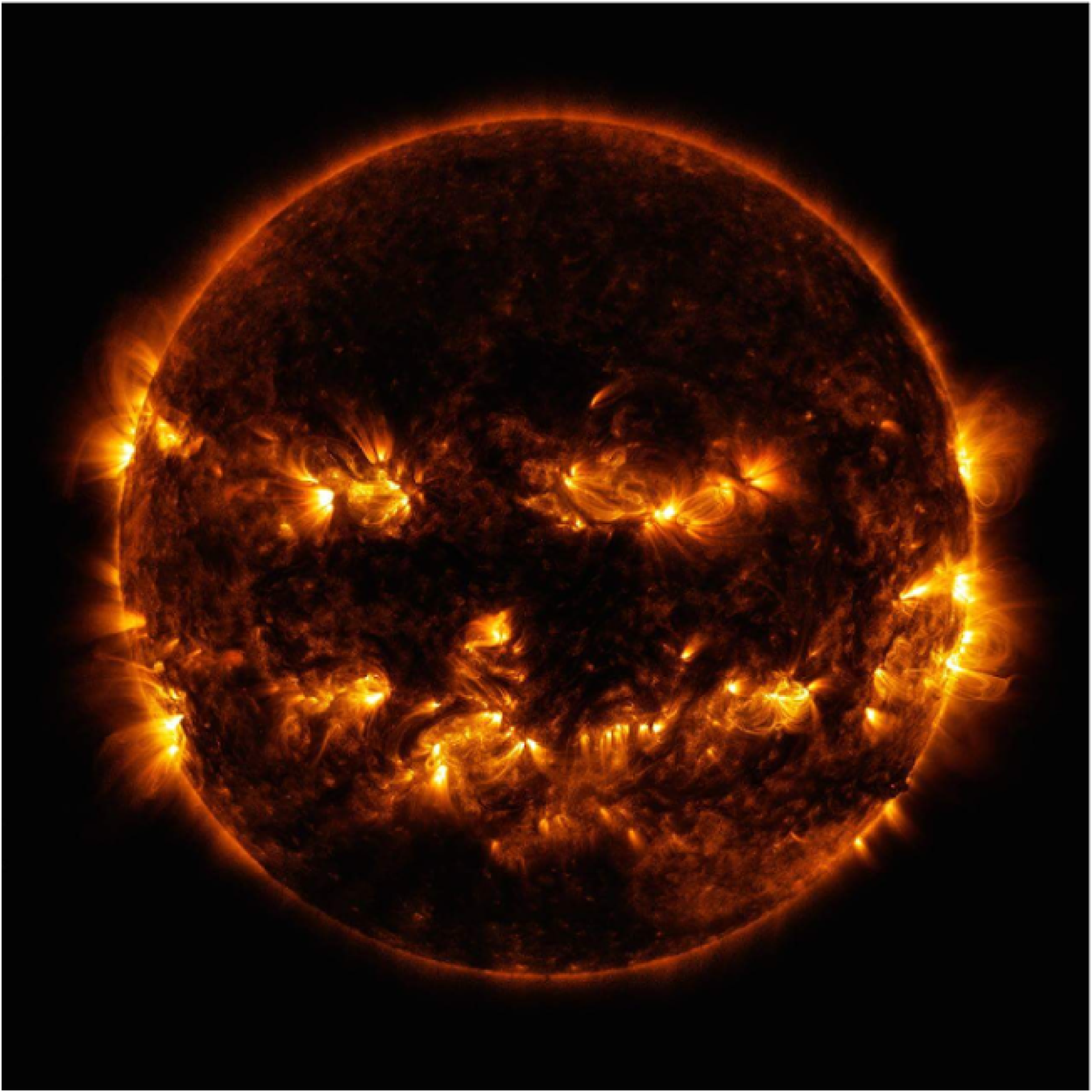}}
\resizebox{0.45\columnwidth}{!}{\%
\includegraphics{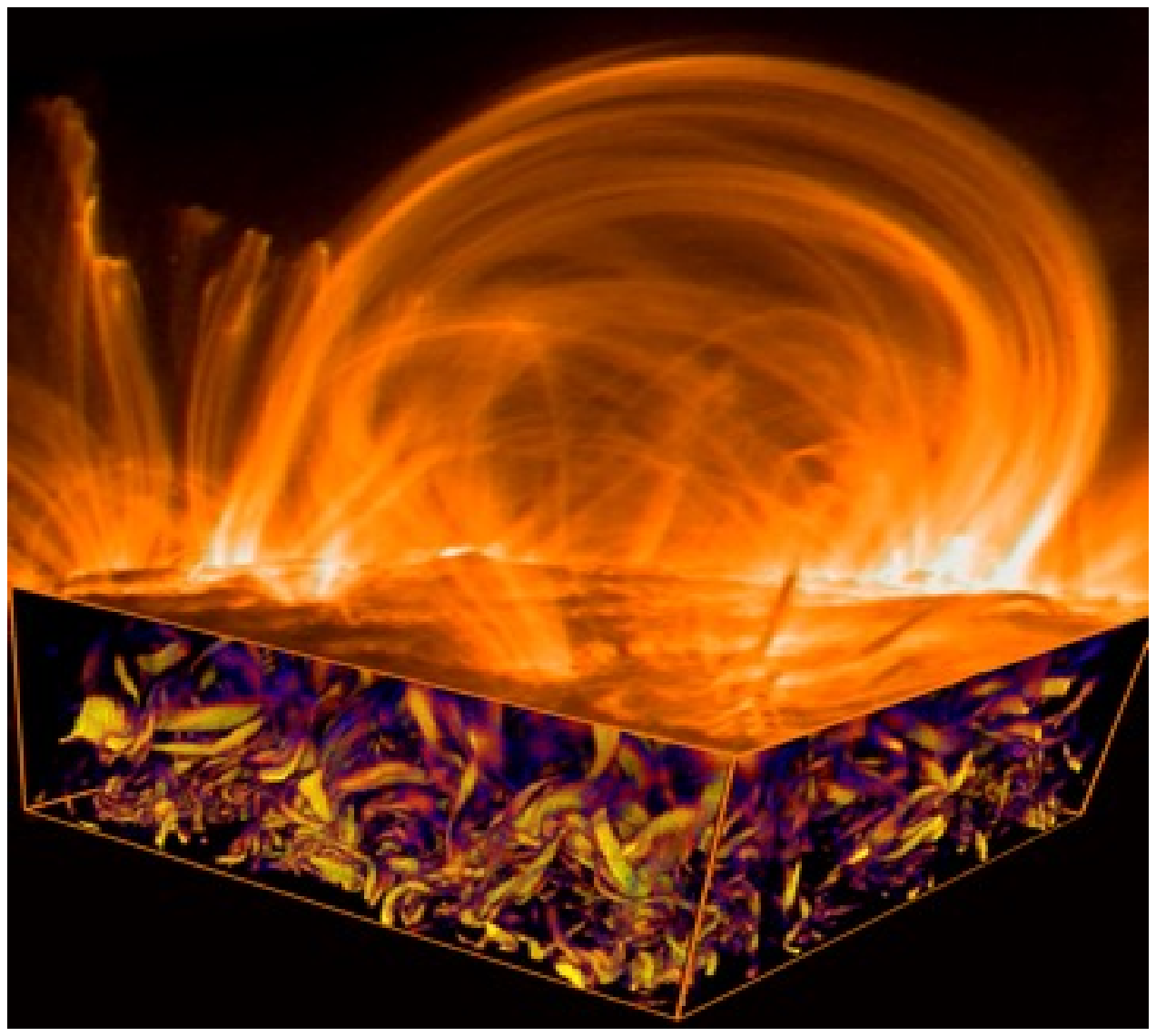}}
\caption{(a) A Full disk picture of the Sun with a dozen active regions over a five-day period (May 14-18, 2015), (b) The turbulent convection zone generates new magnetic flux. Magnetic buoyancy forces the magnetic field towards the solar surface, therefore  it acts as the driver for the emerged magnetic topology. The nonlinear coupling of two dynamical systems is behind many well known astrophysical phenomena. \label{AR5}
}
\end{figure}

ARs are non linear and open dynamical astrophysical systems, where the turbulent driver (convection zone) forces the magnetic filed topology constantly away from the equilibrium since new magnetic threads or loops {\bf emerge} from the
convection zone and are subject to surface diffusive
{\bf random motion}, which leads to magnetic shears and cancellation of magnetic
energy when they collide with other magnetic structures and form current sheets (see Fig. \ref{AR5}).   Since the magnetic
Reynolds number is very large in the
solar corona, MHD theory states that {\bf magnetic energy
can only be released in localized regions where the magnetic field generally creates very steep local gradients with the adjusted strong-field regions, i.e. in unstable current sheets (UCS)}.
Observations of the solar X-ray corona have been reported in the literature
since the early seventies. In the early eighties several authors
showed that the peak-luminosity distribution of flares displays a
well-defined, extended power law with an index $-1.8\pm 0.05$
(see \cite{Lin84,Cro92}).
\begin{figure}[ht]
\centering
\resizebox{0.45\columnwidth}{!}{
\includegraphics{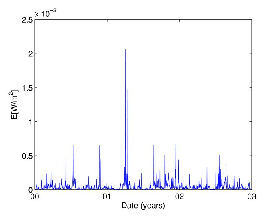}}
\resizebox{0.50\columnwidth}{!}{\includegraphics{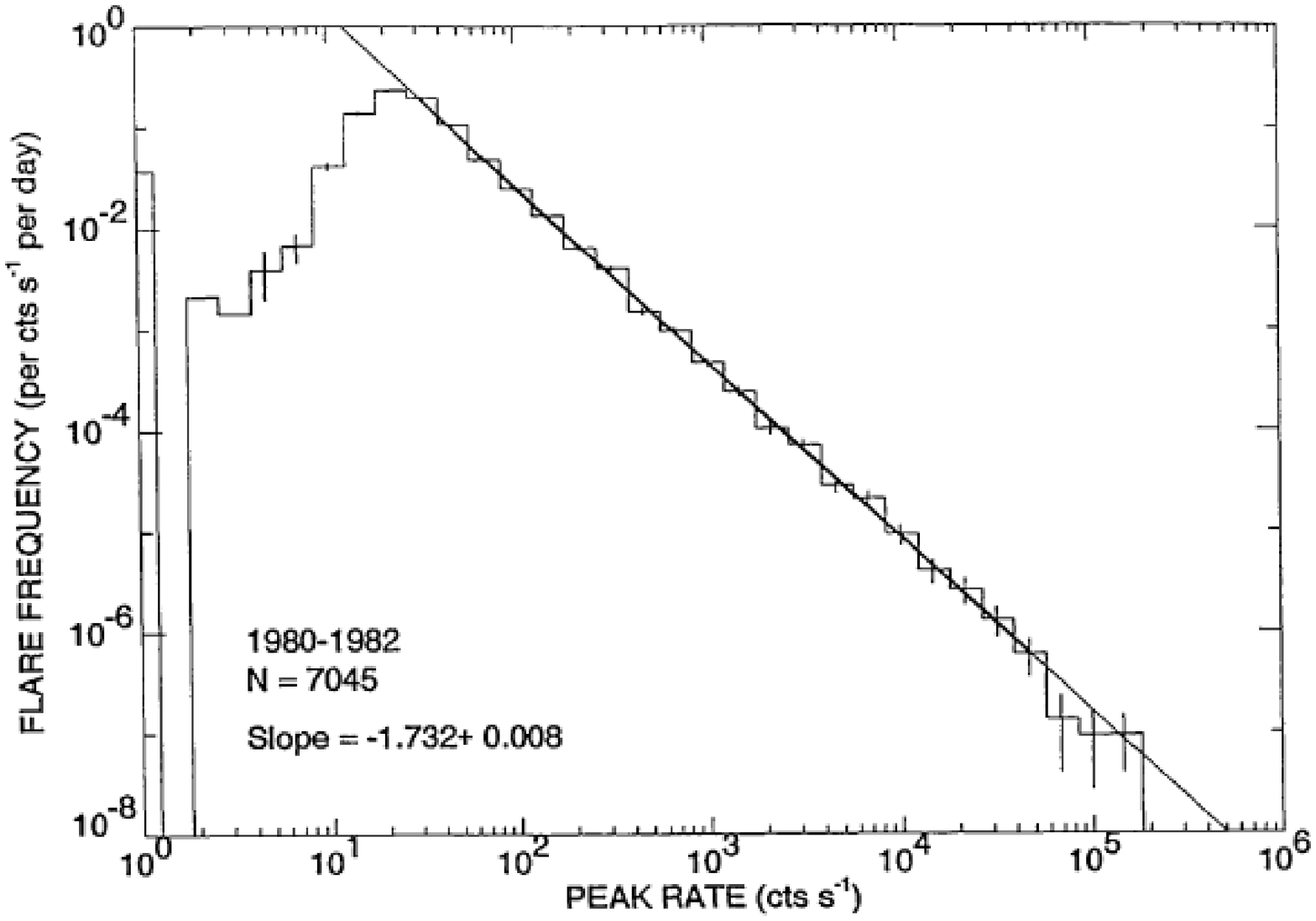}}
\caption{(a) A typical time series of solar flares, (b) The frequency distribution of the peak count rate}
\label{fig:2}       
\end{figure}
Deviations from the power law
behaviour appear in the lowest energies. It has been
pointed out that these deviations  are due to instrumental limitations.

\subsection{Non Linear Magnetic extrapolation of the AR and the formation of magnetic discontinuities}
Numerous articles (see recent reviews \cite{Dem,Longc}) are devoted
to the analysis of magnetic topologies which can host UCSs.
The main trend of current research in this area is to find ways to
realistically reconstruct the 3-D magnetic field topology in the corona
based on the available magnetograms and large-scale plasma motions
at the photosphere.
A realistic magnetic field incorporates many ``poles and sources"
\cite{Longc} and naturally has a relatively large number of UCSs. We
feel that simple representation of the UCS's inside the 3D AR  (dipoles, quadrupoles, symmetric magnetic arcades), while mathematically appealing  \cite{aulanie1} cannot be realistic since
such simple topologies are broken by the photospheric driver, for example due to large-scale sub-Alfv\'enic
photospheric motions or the emergence of new magnetic flux that disturbs the
corona.
All these constraints restrict our ability to reconstruct fully the
dynamically evolving magnetic field of an active region (and it is not clear
that an exact reconstruction will ever be possible).

Dimitropoulou et al. \cite{Dim09,Dim11,Dim13} and Toutountzi et al. \cite{tout} used a model for the 3D reconstruction of the magnetic field in the corona under the force-free assumption. Hence, meaning that the electrical currents flow strictly along the magnetic field lines together with the absence of magnetic monopoles in the bounded volume. In vector notation:

\begin{equation} \label{FF}
\nabla\times\mathbf{B} =\alpha\,\mathbf{B}\ ; \ \  \nabla\cdot\mathbf{B} =0
\end{equation}
\\
where $\alpha$ is the force-free parameter, which in general is a function of position but is conserved along each field line. This is the case of Non-Linear Force-Free (NLFF) fields, which has generally advanced our understanding of the overall morphology of ARs  (see \cite{wiegsakurai} and references therein). A special case is the linear force-free field, in which $\alpha$ is assumed to be constant, but when the extrapolated magnetic field lines are compared to structures from EUV images \cite{wiegsakurai}, it is seen to fail to recover the overall magnetic field topology.
For our investigations we used an optimization technique \cite{wieg04} for computing the NLFF field in the corona. Using appropriate boundary conditions, this numerical method yields a NLFF field solution by minimizing a penalty function, $L$, in the computational volume, $V$, as
\begin{equation}
L= \int \limits_V \, w(x,y,z)[B^{-2}|(\nabla\times \mathbf{B})\times \mathbf{B}|^{2}+|\nabla \cdot \mathbf{B}|^{2}]\, \mathrm{d} V
\end{equation}\\
where $w(x,y,z)$ is a scalar function with a value of 1 in the physical domain of the volume that drops smoothly to zero when approaching the top and lateral boundaries. When $L=0$, both the Lorentz force is zero and the solenoidal condition is satisfied in the entire computational volume, which then contains the NLFF field.

\begin{figure}[ht]
\centering
\resizebox{0.45\columnwidth}{!}{
\includegraphics{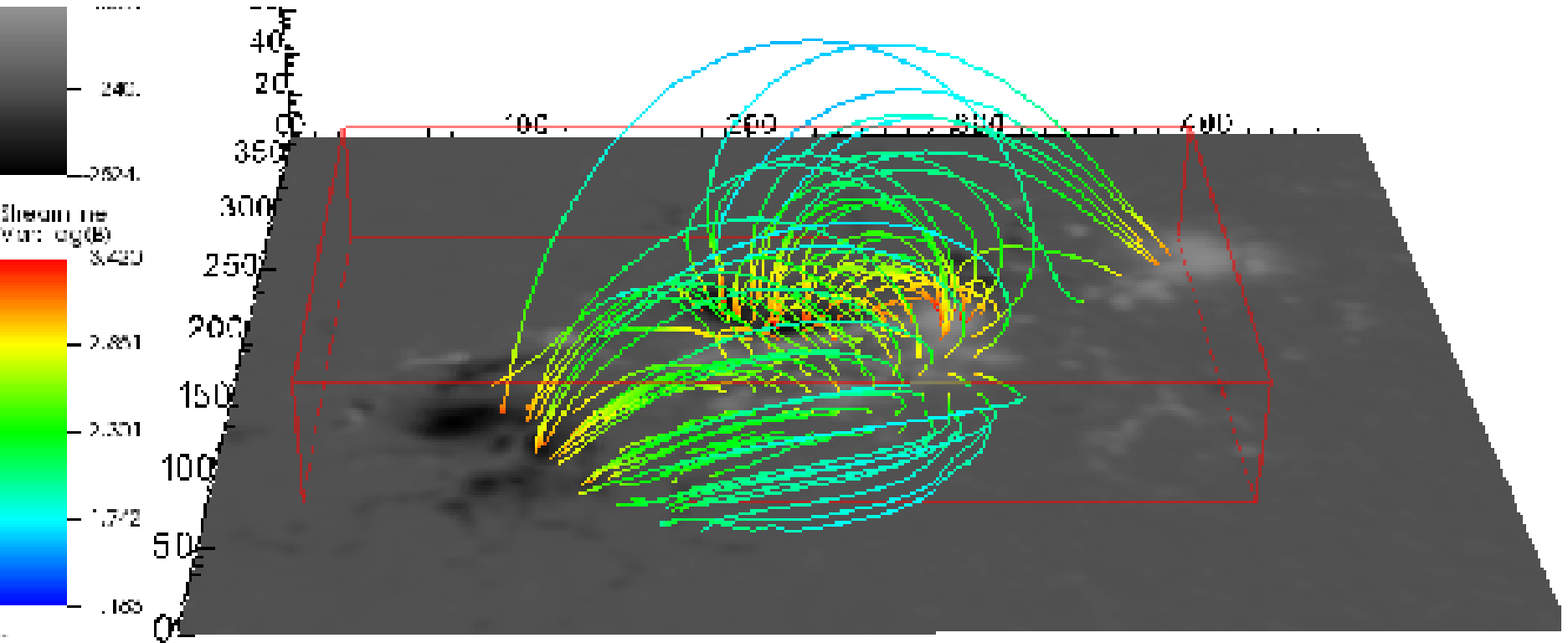}}
\resizebox{0.45\columnwidth}{!}{
\includegraphics{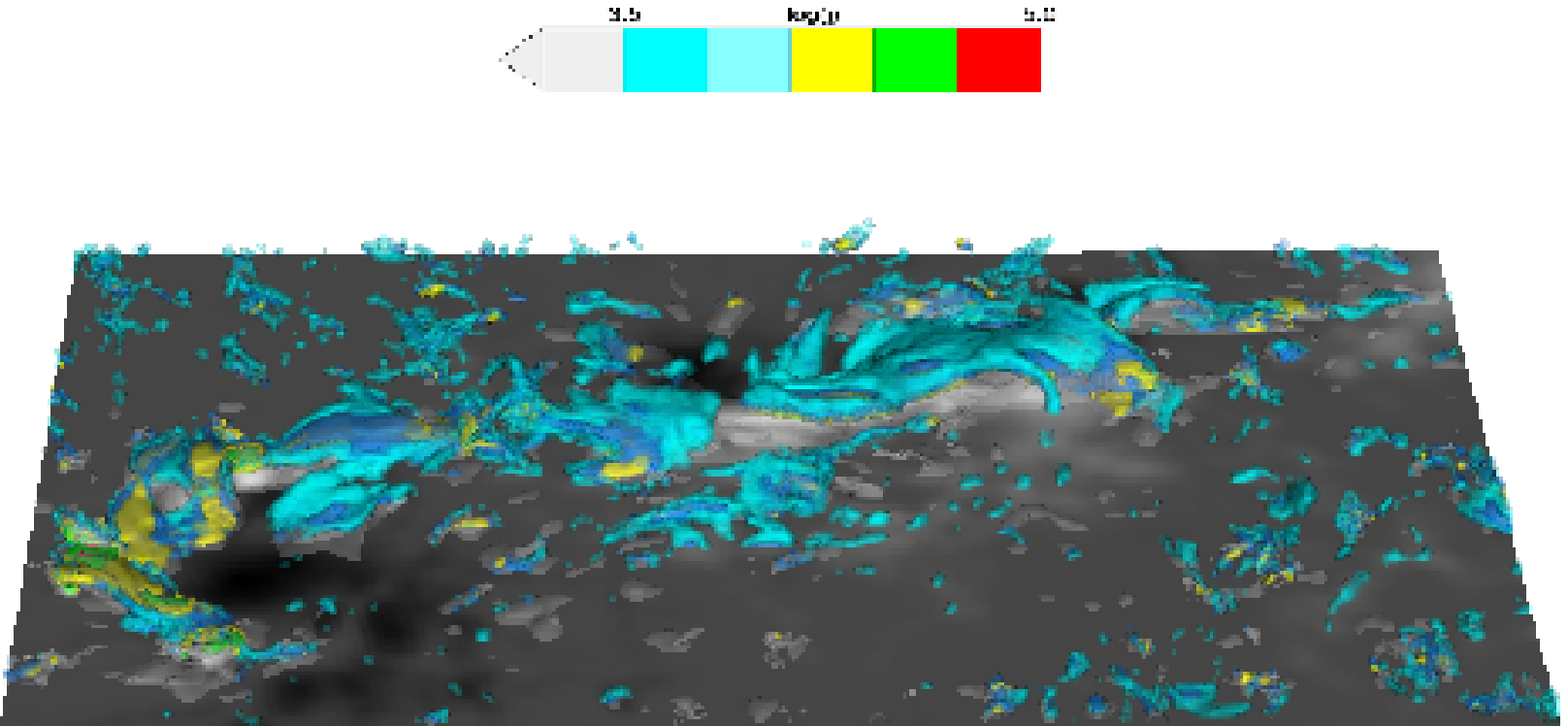}}
\caption{(a)Original magnetogram for the eruptive NOAA AR 11158 on February 14, 2011 at 21:58 UT together with the NLFF field lines. (b) Spontaneous formation of UCS in the low corona due to the twist of magnetic fields in a complex magnetic topology \cite{tout}  \label{figdata}}

\end{figure}

 An example of the resulting 3D NLFF field is shown in Fig.~\ref{figdata}a. In \ref{figdata}b the isosurfaces of the current density ( $\nabla \times \vec{B}$  correspond to the red cube of the \ref{figdata}a. It is important to  notice that the UCS are concentrated in the low corona in the quiet AR.

 \vspace{2.5 cm}

Dimitropoulou at al. \cite{Dim09} studied the corelation of the fractal structures in the photosphere and the coronal magnetic field. They conclude that there is no correlation between the 2D fractal dimension in the photosphere and the 3D fractal dimension in the NLFF reconstructed magnetic topology.
Photospheric turbulence remains the driver
for the coronal instabilities, but the strong nonlinearity of the
system in the lower coronal layers destroys any kind of direct
relation between the photospheric structures and their coronal
counterparts. The photospheric driver forces the system to accumulate
a large number of magnetic discontinuities that store
enough energy to explain the statistical properties of the solar
activity in case of release. These discontinuities form patterns
that do not follow the morphological properties of the photospheric
magnetic flux concentrations, but have a strong impact
on the expected dynamical activity of the system as we will see next.

The spatial distribution of the
UCS with height shows that $80\%$ of the magnetic discontinuities
are accumulated in the lower corona (within 20 Mm
from the photosphere). The system is evidently highly unstable
at these heights, due to  processes that are clearly nonlinear.
It is this strong nonlinearity at lower layers that does not allow
the corona to respond proportionally to changes imposed by the photospheric driver.

\subsection{The local and global evolution of UCS}

\subsubsection{Local evolution of an isolated or a small number of UCS}
Research on  reconnecting magnetic fields has  undergone a dramatic evolution recently due mostly to the development of the numerical simulation techniques.  Long current sheets or multiple interactig current sheets will form on a short time scale  a new collection of current sheets which are the results of the current fragmentation\cite{Drake06,marco,marco1,Hoshino,Gordo11,Kowal11,Barta,Baumann12,Nishi13,Zhd}, (see also recent reviews \cite{vla08,Cargill12,Lazarian12}).  On the other hand, Alfven waves and large scale disturbances travelling along complex magnetic topologies will drive magnetic discontinuities by reinforcing existing current sheets or form new unstable current sheets (see \cite{Matt86,Biskam89,Ambrosiano88,Georgoulis98,Dmitruk03,Dmitr2,Dahl12,Eunaudi,Arzner04,Buc,Arzner06,Turkmani06} ).  The interplay between turbulence and magnetic reconnection has been studied recently in several publications \cite{Laza99,Wan,Karimabadi03} and the statistical properties of the magnetic reconnection sites \cite{Georgoulis98,Dmitr98,Servi09,Servi10}

\begin{figure}[ht]
\centering
\resizebox{0.75\columnwidth}{!}{%
\includegraphics{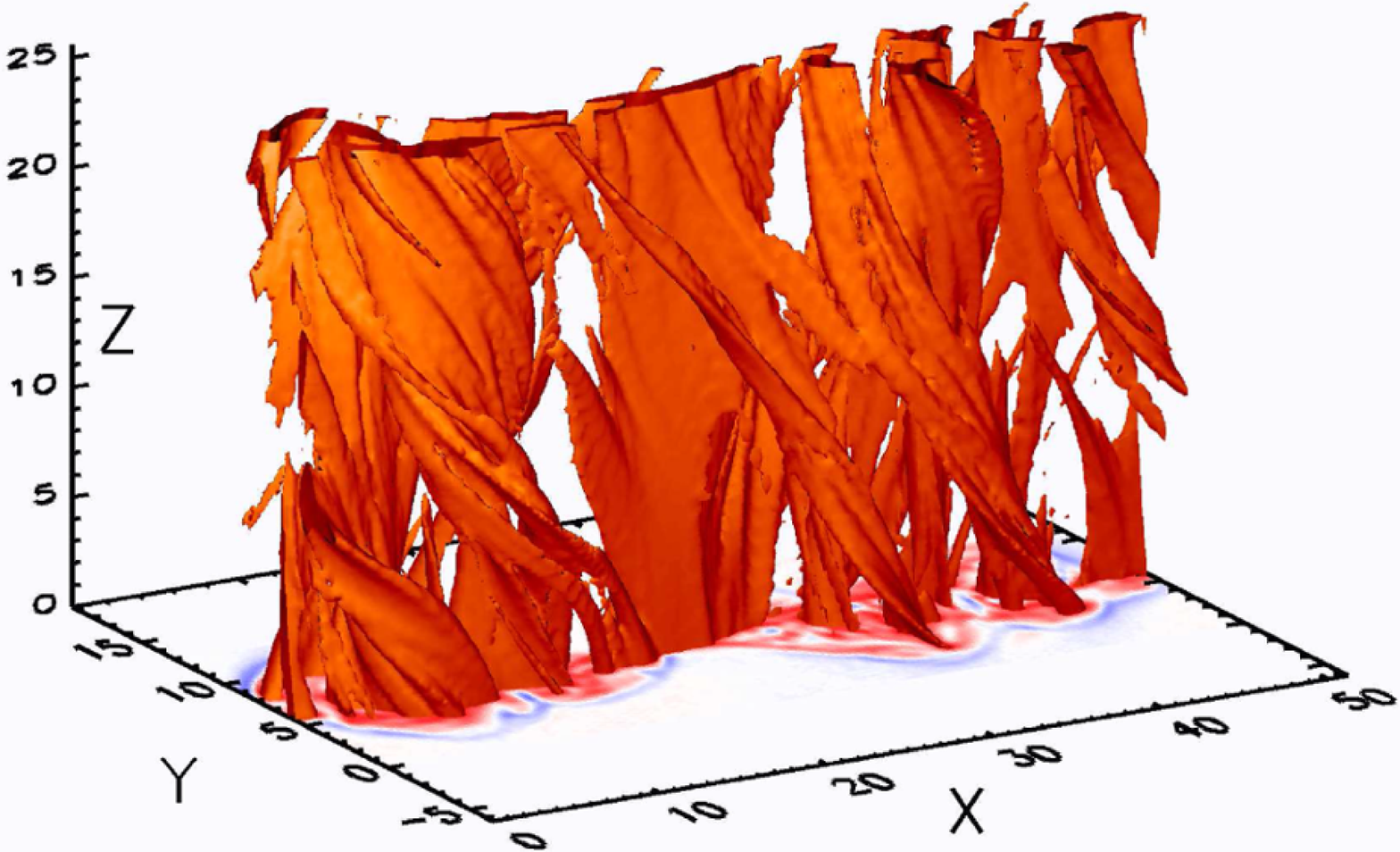}}\\
\resizebox{0.4\columnwidth}{!}{%
\includegraphics{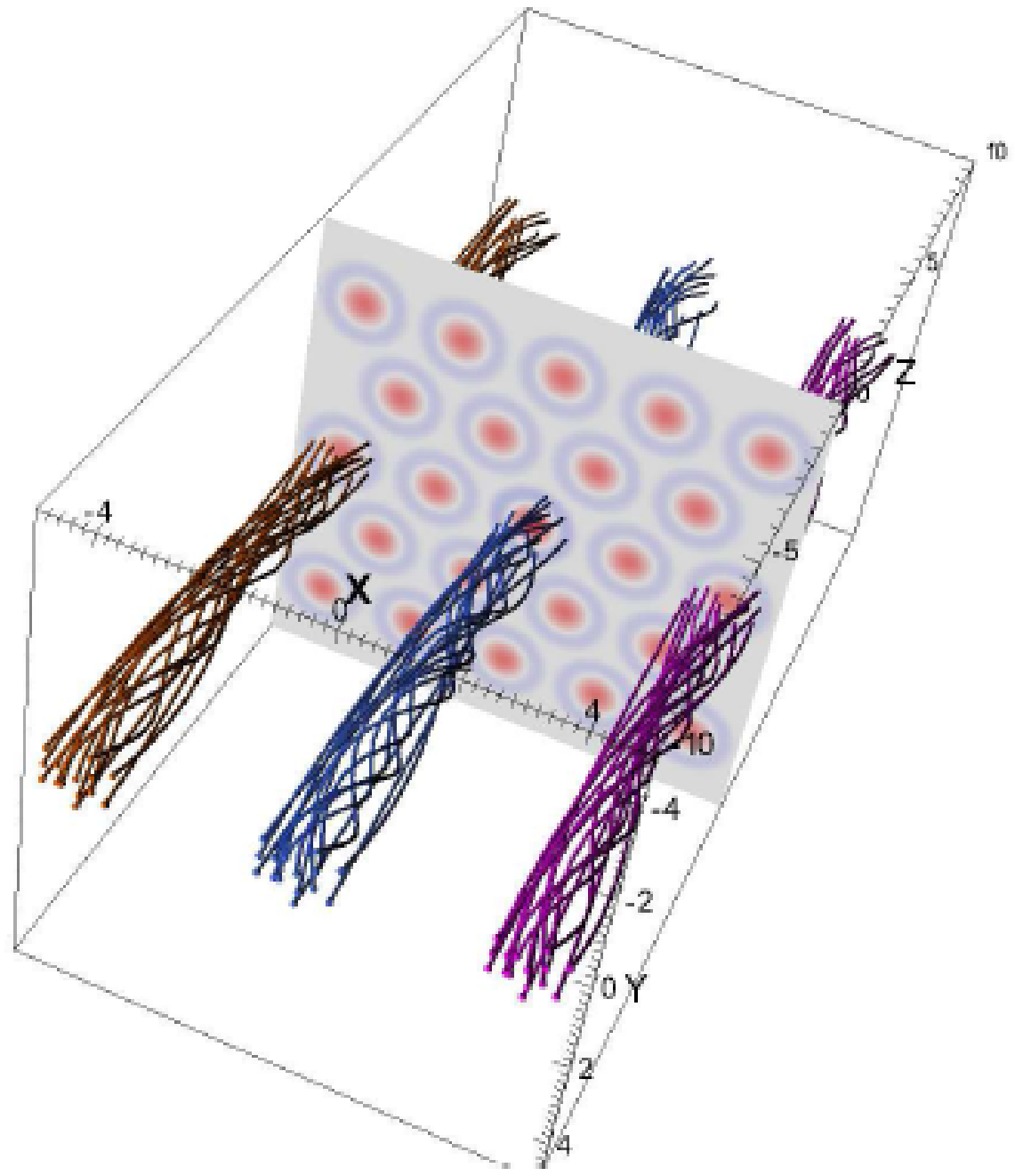}}
\resizebox{0.4\columnwidth}{!}{%
\includegraphics{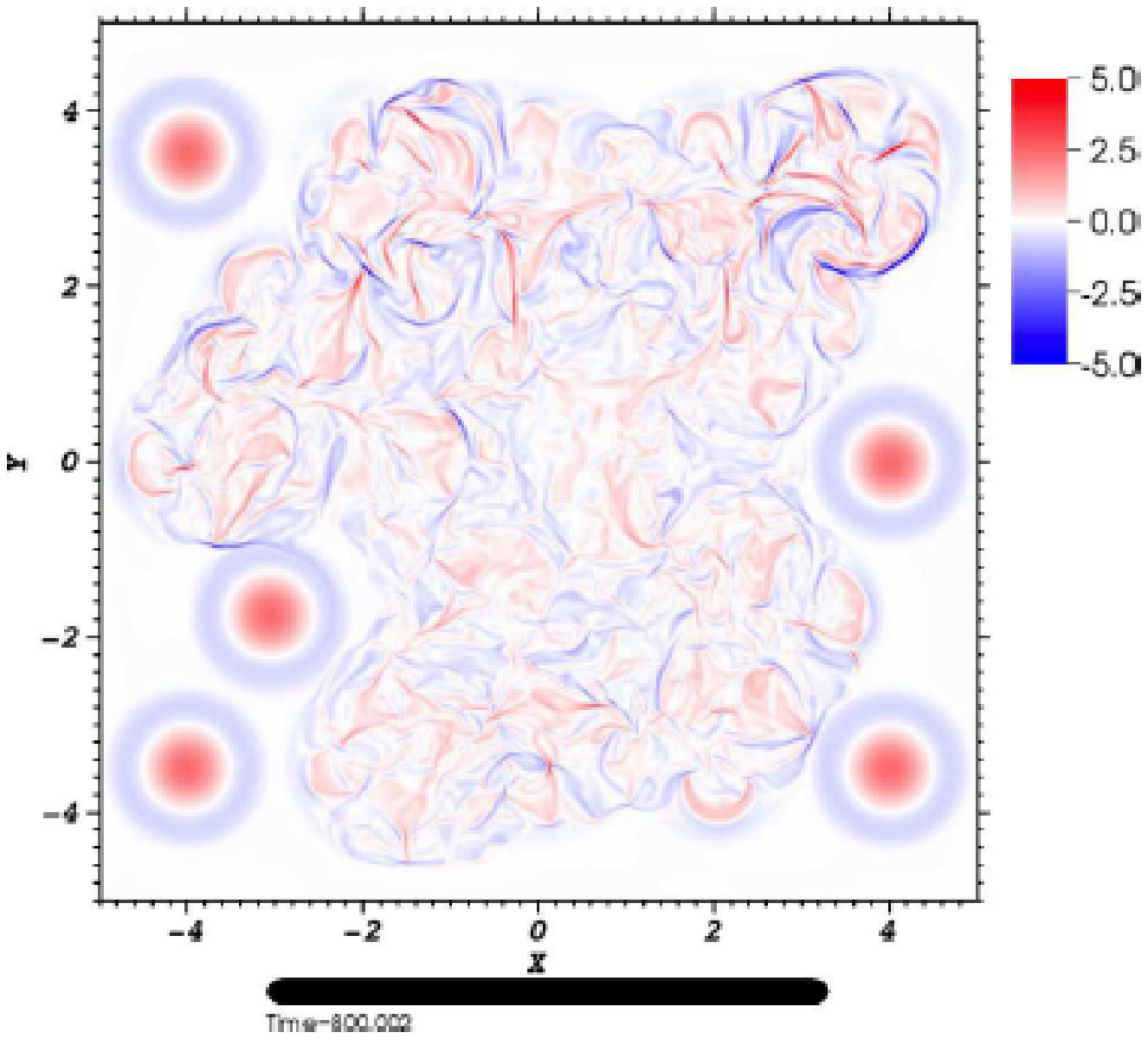}}
\caption{(a) surface of $J_{z}$ at $tÃÂÃÂÃÂÃÂÃÂÃÂÃÂÃÂ©_{ci}= 50$. The isosurface level is $60 \%$ of the maximum current density (a 2D slice of the same quantity is shown on the bottom) \cite{Dahlin} (b) Twenty three threads are used in the avalanche simulation. The
twisted field lines outline just three of the threads. The contours of the axial
current density are shown in the mid-plane. The gaps between the threads are
filled with a uniform axial field. (c) Contours of the axial current at the mid plane $(z = 0)$ during the later stages when the avalanche ocurred.  Here the background resistivity is
zero. Red corresponds to positive current, blue to negative and white to zero. \cite{Hood15}   \label{sfrag}}

\end{figure}

The most important point reported in many studies is the fragmentation of the UCS and the redistribution of magnetic flux   \cite{marco1,Dahlin,Hoshino,Zhd} (see Fig. \ref{sfrag}a)  which is closely related with one of the basic rules of Self-Organised Criticality reported in the next section. Hood at al. \cite{Hood15} moved one step further and demonstrated for the first time how an MHD avalanche might occur in a multi thread coronal loop system (see Fig. \ref{sfrag}b). They showed how once one stable thread is disrupted, it coalesces with neighboring threads and this process continues disrupting more and more threads and an avalanche will occur \cite{Hood15}.

\subsubsection{Global evolution of UCS, Self-Organized Criticality and Turbulent Reconection}
The existence of power laws in the frequency distribution of the
explosive activity (see Fig. \ref{fig:2})  may suggest that explosions are a self-organization
phenomenon in the AR. Lu and Hamilton \cite{Lu91} were  the first to
realize that ARs may be in the Self-Organized Critical
state and proposed that explosions  ultimately are caused by small magnetic
perturbations $(\delta B)$ ({\bf Loading}) which gradually force the CS become UCS when a {\bf critical threshold} is passed. The local fragmentation of the UCS causes a re-organization of
the unstable magnetic topology which may cause
{\bf avalanches}   of
all sizes (nano-flares, micro-flares, flares) (the basic ideas of SOC were initially proposed by Bak et al. \cite{Bak87} twenty five years ago). This model opened
the way for a series of similar models developed during the  last twenty five years (see reviews by \cite{Jen98,Cha02,Ascwa,Ascwa15}).

There are many ways to develop  CA models to represent  SOC \cite{Ascwa}, one of them based its rules on the MHD equations \cite{Isl00,Isl01}. The proposed  set-up can be superimposed
onto each classical solar flare CA model, and which makes the
latter interpretable in a MHD-consistent way (by {\it classical}
CA models we mean the models of \cite{Lu91} (LH91) and
their modifications, which are based on the sand-pile analogy \cite{Vla95,Geo96,Geo98,Gal96}).
The set-up thus specifies the physical interpretation of the
grid-variables and allows the derivation of quantities such as
currents etc. It does not interfere with the dynamics of the CA
(unless wished): loading, redistributing (bursting), and the
appearance of avalanches and Self-Organized Criticality (SOC), if the latter are implied by the evolution rules, remain unchanged.
The result is therefore still a CA model, with all the advantages
of CA, namely that they are fast, that they model large spatial
regions (and large events), and therewith that they yield good
statistics. Since the set-up introduces all the relevant physical
variables into the context of the CA models, it automatically
leads to a better physical understanding of the CA models. It
reveals which relevant plasma processes and in what form are
actually implemented, and what the global flare scenario is the CA
models imply. All this was more or less hidden so far in the
abstract evolution rules. It leads also to the possibility to
change the CA models (the rules) at the guide-line of MHD, if this
should become desirable. Not least, the set-up opens a way for
further comparison of the CA models to observations.

The specifications the set-up meets are : The vector $\vec
A_{ijk}$ at the grid sites $\vec x_{ijk}$ denotes the local
vector-field, $\vec A(\vec x_{ijk})$. Note that this was not
specified in the classical CA models. Lu et al. \cite{Lu93} for instance
discuss this point: it might also have been thought of as a mean
local field, i.e. the average over an elementary cell in the grid.

Guided by the idea that we want to assure $\nabla \cdot \vec B = 0$ for
the magnetic field $\vec B$, which is most easily achieved by
having the vector-potential $\vec A$ as the primary variable and
letting $\vec B$ be the corresponding derivative of $\vec A$
($\vec B = \nabla \times \vec A$), we furthermore assume that the
grid variable $\vec A$ of the CA model is identical with the
vector-potential.

The remaining and actually most basic problem then is to find an
adequate way to calculate derivatives in the grid. In general, CA
models assume that the grid-spacing is finite, which also holds
for the CA model of \cite{Lu91} (as shown in detail by
\cite{Isl98}, so that the most straightforward way of replacing
differential expressions with difference expressions is not
adequate. Consequently, one has to find a way of continuing the
vector-field into the space in-between the grid-sites, which will
allow to calculate derivatives. For this purpose we use spline interpolation, where the 3D interpolation is
performed as three subsequent 1D interpolations in the three
spatial directions (\cite{Pre92}). For the 1D splines, we assume
natural boundaries (the second derivatives are zero at the
boundaries).

With the help of this interpolation, the magnetic field $\vec B$
and the current $\vec J$ are calculated as derivatives of $\vec
A$, according to the MHD prescription:
\begin{equation}
\vec B = \nabla \times \vec A,
\end{equation}
\begin{equation}
\vec J = {c \over 4\pi} \, \nabla \times \vec B.
\end{equation}

According to MHD, the electric field is given by Ohm's law, $\vec
E = \eta \vec J - {1\over c} \vec v \times \vec B$, with $\eta$
the diffusivity and $\vec v$ the fluid velocity. Since the
classical CA models use no velocity-field, our set-up can yield
only the resistive part,
\begin{equation}\label{res}
\vec E = \eta \vec J.
\end{equation}
In applications such as to solar explosions, where the interest is in
current dissipation events, i.e.\ in events where $\eta$ and $\vec
J$ are strongly increased, Eq. \ref{res} can be expected to be a
good approximation to the electric field. Theoretically, the
convective term in Ohm's law would in general yield just a
low-intensity, background electric field.

Eq. \ref{res} needs to be supplemented with a specification of the
diffusivity $\eta$: \cite{Isl98} have shown that in the classical
CA models the diffusivity adopts the values $\eta =1$ at the
unstable (bursting) sites, and $\eta =0$ everywhere else. This
specifies Eq.\ref{res} completely.
This set-up of Isliker et al. \cite{Isl00,Isl01} for classical solar flare CA
models yields, among others, consistency with Maxwell's equations
(e.g.\ divergence-free magnetic field), and availability of
secondary variables such as currents and electric fields in
accordance with MHD.

The main aim in \cite{Isl00,Isl01} with the introduced set-up was
to demonstrate that the set-up truly extends the classical CA
models and makes them richer in the sense that they contain much
more information, now. The main features we revealed about the CA
models, extended with this set-up, are:

\noindent {\bf 1.\ Large-scale organization of the
vector-potential and the magnetic field:}
The field topology during SOC state is bound to characteristic
large-scale structures which span the whole grid, very pronounced
for the primary grid variable, the vector-potential, but also for
the magnetic field . Bursts and flares are just slight
disturbances propagating  over the large-scale structures, which
are always maintained, also in the largest events.

\begin{figure}[ht]
\centering
\resizebox{0.40\columnwidth}{!}{%
\includegraphics{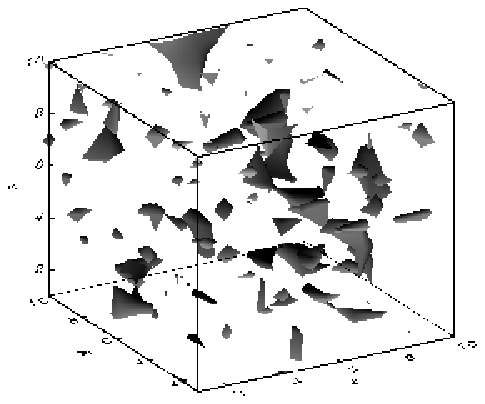}}
\resizebox{0.40\columnwidth}{!}{%
\includegraphics{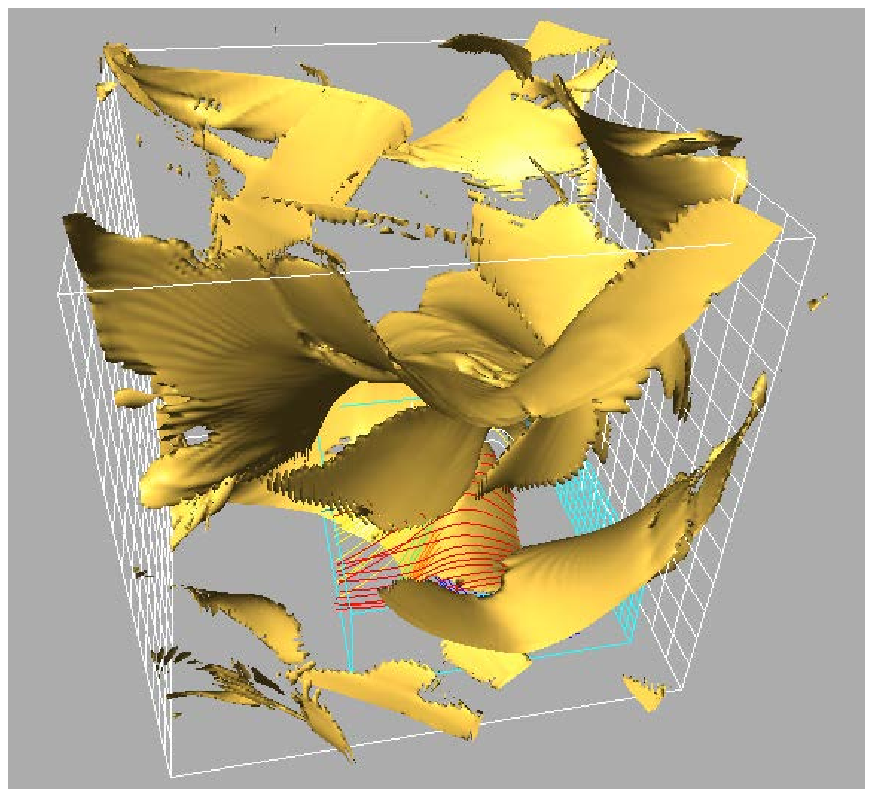}}
\caption{(a) Three dimensional isosurfaces of electric current
density derived from the CA model (\cite{Isl01}) (b) Isosurfaces
of electric current density, from 3D MHD experiments with boundary
driven magnetic dissipation (\cite{NorGal96,Galsgaard})}
\label{CS1}
\end{figure}

\noindent {\bf 2.\ Increased current at unstable grid-sites:}
Unstable sites are characterized by an enhanced current, which is
reduced after a burst has taken place, as a result of which the
current at a grid-site in the neighbourhood may be increased.

\noindent {\bf 3.\ Availability of the electric field:} The
electric field is calculated with the resistive part of Ohm's law,
which can be expected to be a good approximation in applications
where the interest is in current-dissipation events, e.g.\ in the
case of solar flares.



\noindent {\bf 4.\ Energy release in terms of Ohmic dissipation:}
We replaced the some-what {\it ad hoc formula} in the CA models to
estimate the energy released in a burst
with the expression for Ohmic dissipation in terms of the current.
The distributions yielded in this way are very similar to the ones
based on the ad hoc formula, so that the results of the CA models
remain basically unchanged.

\noindent {\bf 5.\ CA as models for current dissipations:} As a
consequence of point 2 and 4 in this list, and of the fact that
there is an  approximate linear relation between the current and
the stress measure of the CA, we can conclude that the {\it
extended} CA models can be considered as models for energy release
through current dissipation.

It is important to mention here another attempt made by
\cite{NorGal96,Gal96,Galsgaard} to simulate, using a 3-D MHD code, the sporadic
development and evolution of current sheets of all sizes inside
an active region. They used periodic ``y-z'' boundaries and
perfectly conducting rigid x-boundaries with sinusoidal shear with
randomly changing direction and phase, acting on an initial
magnetic field with straight field lines. It is remarkable to note
the appearance of non-steady current surfaces as it is the case
with the CA model presented above (see Fig.~\ref{CS1}).
The
boundary motion used in these studies is still very simple and has
no direct influence on the characteristics of the energy release.
Fragos et al. \cite{Fragos03} used the ``magnetograms" developed with the percolation method and a linear extrapolation to search for the statistical properties of the reconstructed AR and Vlahos and Georgoulis  \cite{vla04} did the same using observations, noticed that  the re-organization of the magnetic fields is a potential way to identify UCS in the coronal part of the AR.
Dimitropoulou et al. \cite{Dim13} use a series of observed magnetogarms to drive the SOC model proposed by Isliker et al. \cite{Isl00,Isl01}. They obtain robust power laws in the distribution functions of the modeled flaring events with scaling law indices that agree well with the observations.
\begin{figure}[ht]
\centering
\resizebox{0.45\columnwidth}{!}{\includegraphics{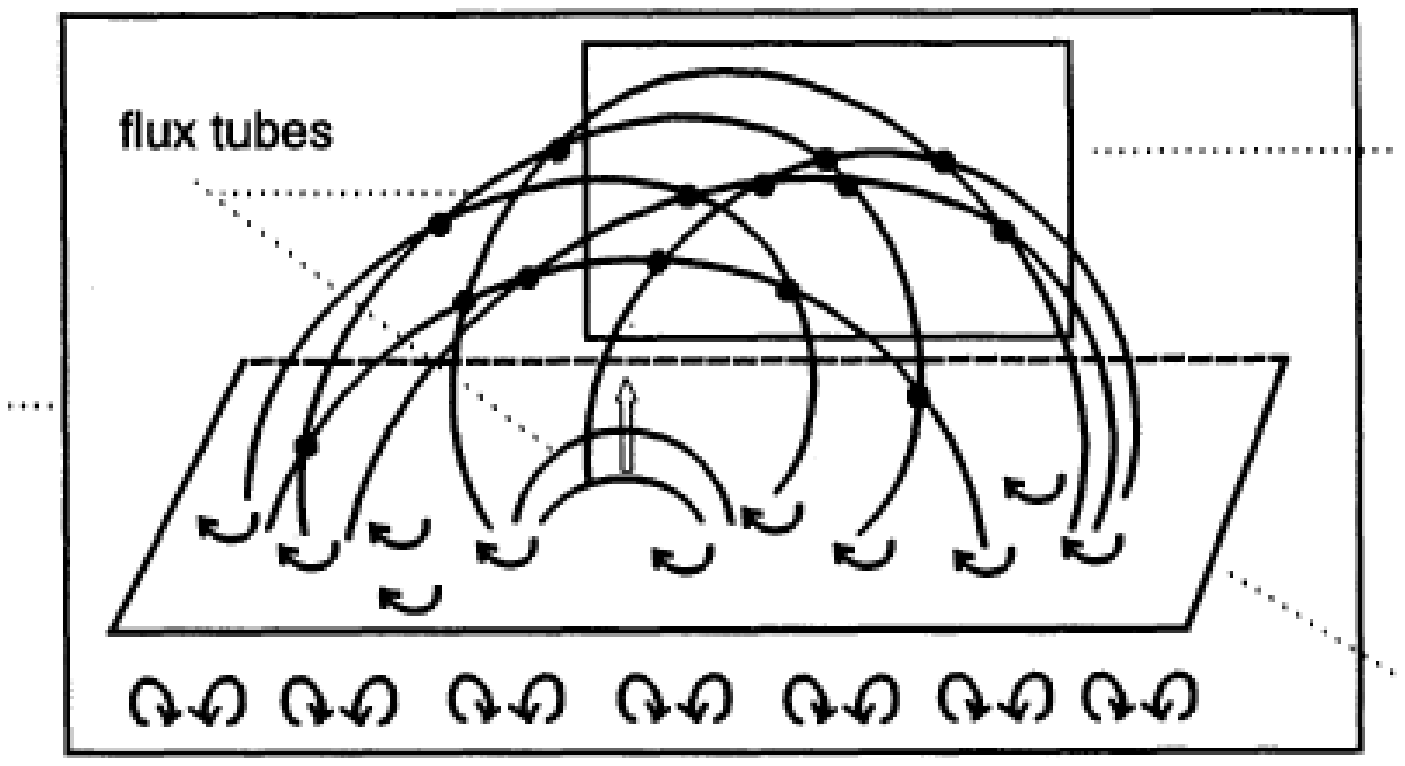}}
\resizebox{0.45\columnwidth}{!}{\includegraphics{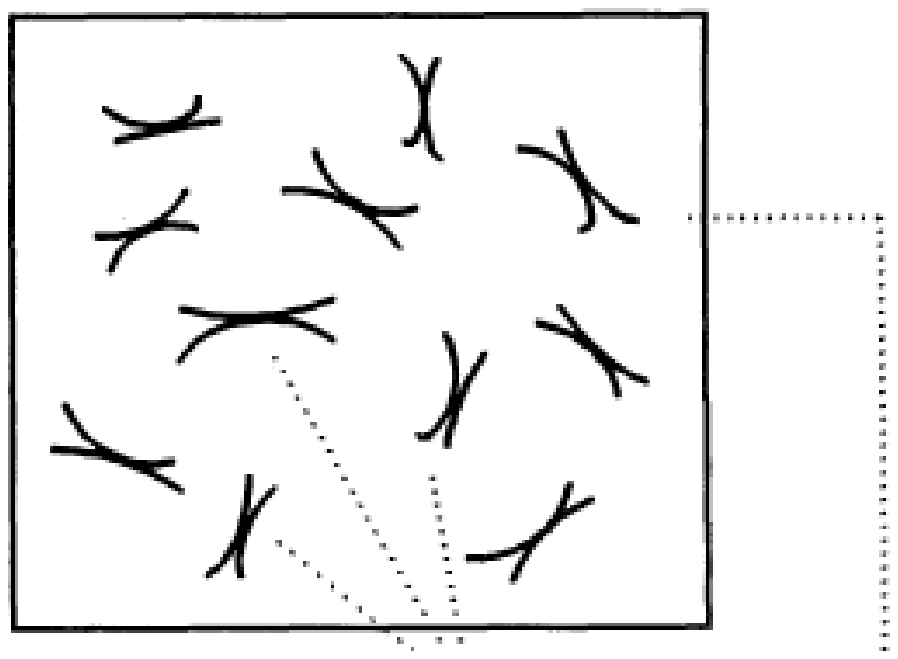}}\\
\resizebox{0.30\columnwidth}{!}{
\includegraphics{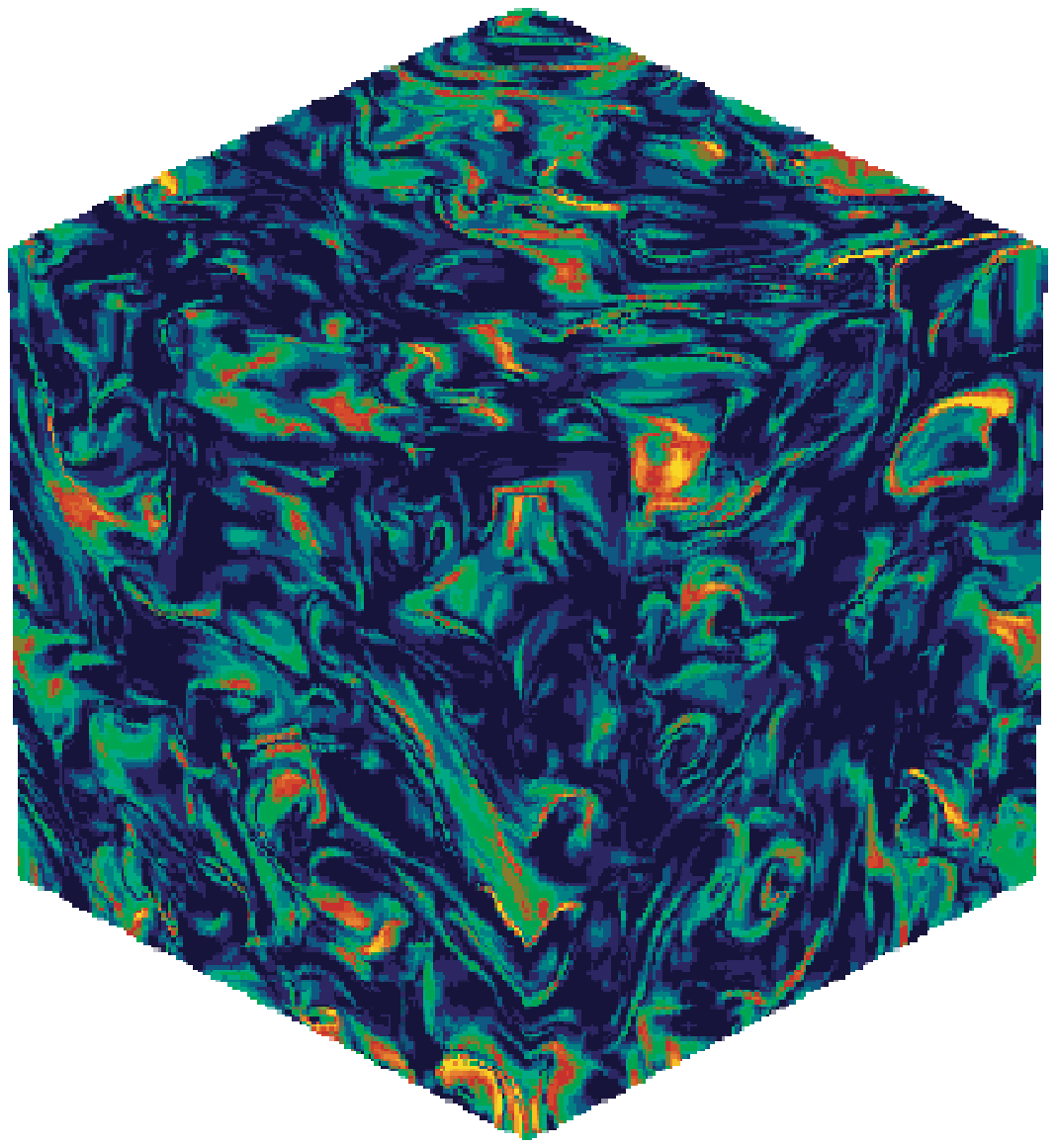}}\resizebox{0.30\columnwidth}{!}{
\includegraphics{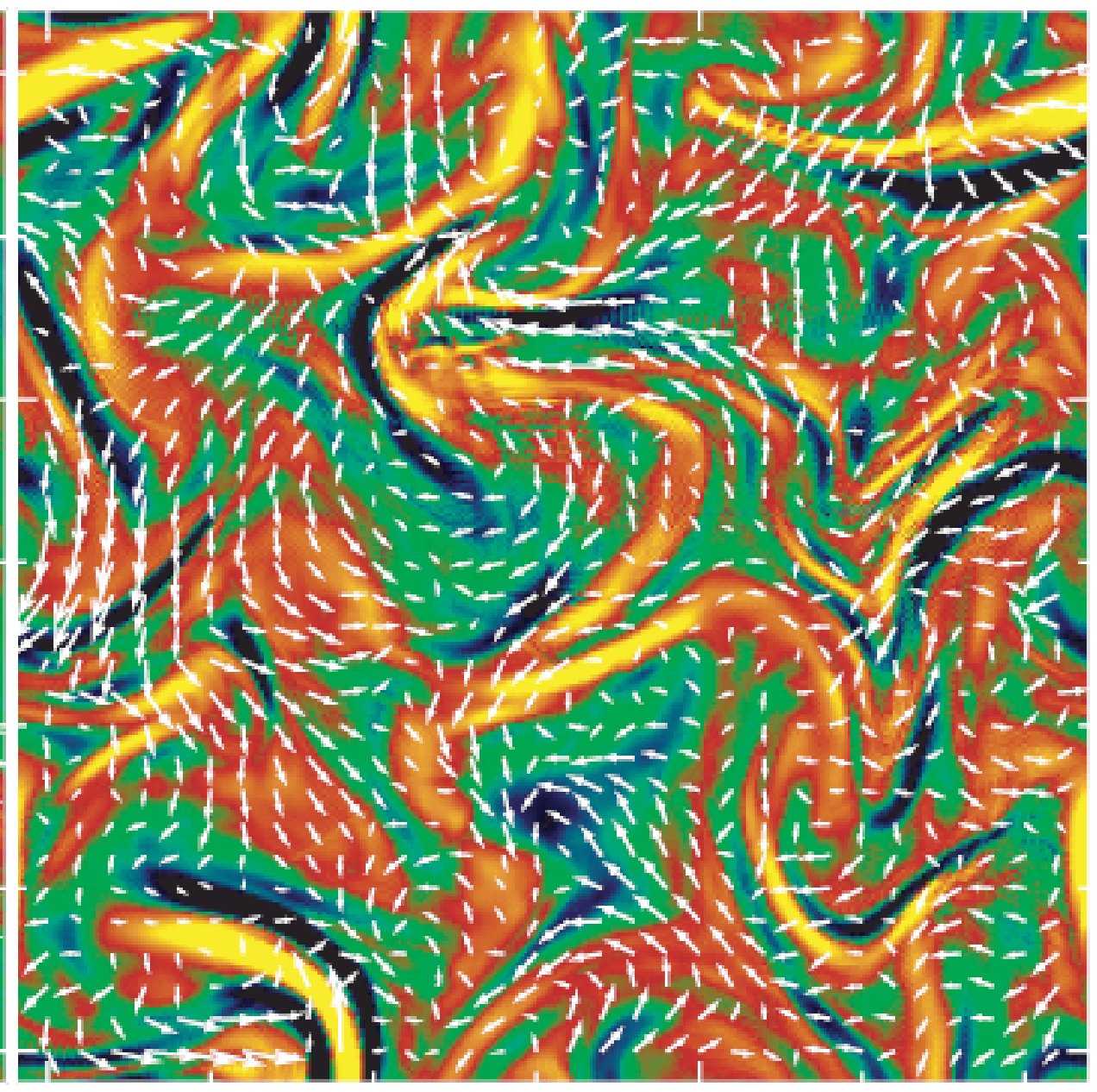}}
\caption{(a) Sketches of UCS developed intermittently at
random positions \cite{Anastasi94} , (b) 3D MHD simulations and visualization of the turbulent electric field $\mid E \mid$ in the simulation box (left). High values are in yellow (light) and low values in blue (dark) and cross section of the current density $J_z$ along the external magnetic field in color tones (right). Yellow (light) is positive $J_z$, blue (dark) is negative, and the superposed arrows represent the velocity field \cite{Dmitr2}}.\label{vl1}
\end{figure}

Models along these lines have been proposed
(\cite{vla84,vla93,vla94,Anastasi94} see also Fig.~\ref{vl1})  in the mid
80's and the beginning of the 90's and remain undeveloped due to
the lack of tools for the global analysis of the active regions
till recently.
The nonlinear coupling of the turbulent convection zone with the AR and the consistency of the results obtained by the SOC theory with those expected by  turbulence has been studied intensively by several authors \cite{Uritsky3,Uritsky2,Uritsky1}. Uritsky et al. \cite{Uritsky3} examined in depth the question of the relation of SOC with turbulence in the solar Corona and agree  with the suggestion made  by Dahlburg et al. \cite{Dahl} that UCS and their \
{\bf fragmentation can serve as the driver for the avalanches in the SOC} scenario. Uritsky \& Devila \cite{Uritsky2} also suggested by studying an AR in a quiescent non-flaring period that (1) there is  formation of non-potential magnetic structures with complex polarity separation lines inside the active region, and (2) there are statistical signatures of canceling bipolar magnetic structures coinciding with flaring activity in the active region. Each of these effects can represent an unstable magnetic configuration acting as an energy source for coronal dissipation and heating.
The development of a parallel use of models based on the Complexity theory and the well established 3D MHD or Kinetic codes is the only way to explore the interplay between global and local scales in turbulent systems.

\section{Anomalous energy transport in turbulently reconnecting Active Regions}
\label{sec:4}
\subsection{Systems far from equilibrium: The rotating annulus}
\begin{figure}[ht!]
\centering
\resizebox{0.35\columnwidth}{!}{  \includegraphics{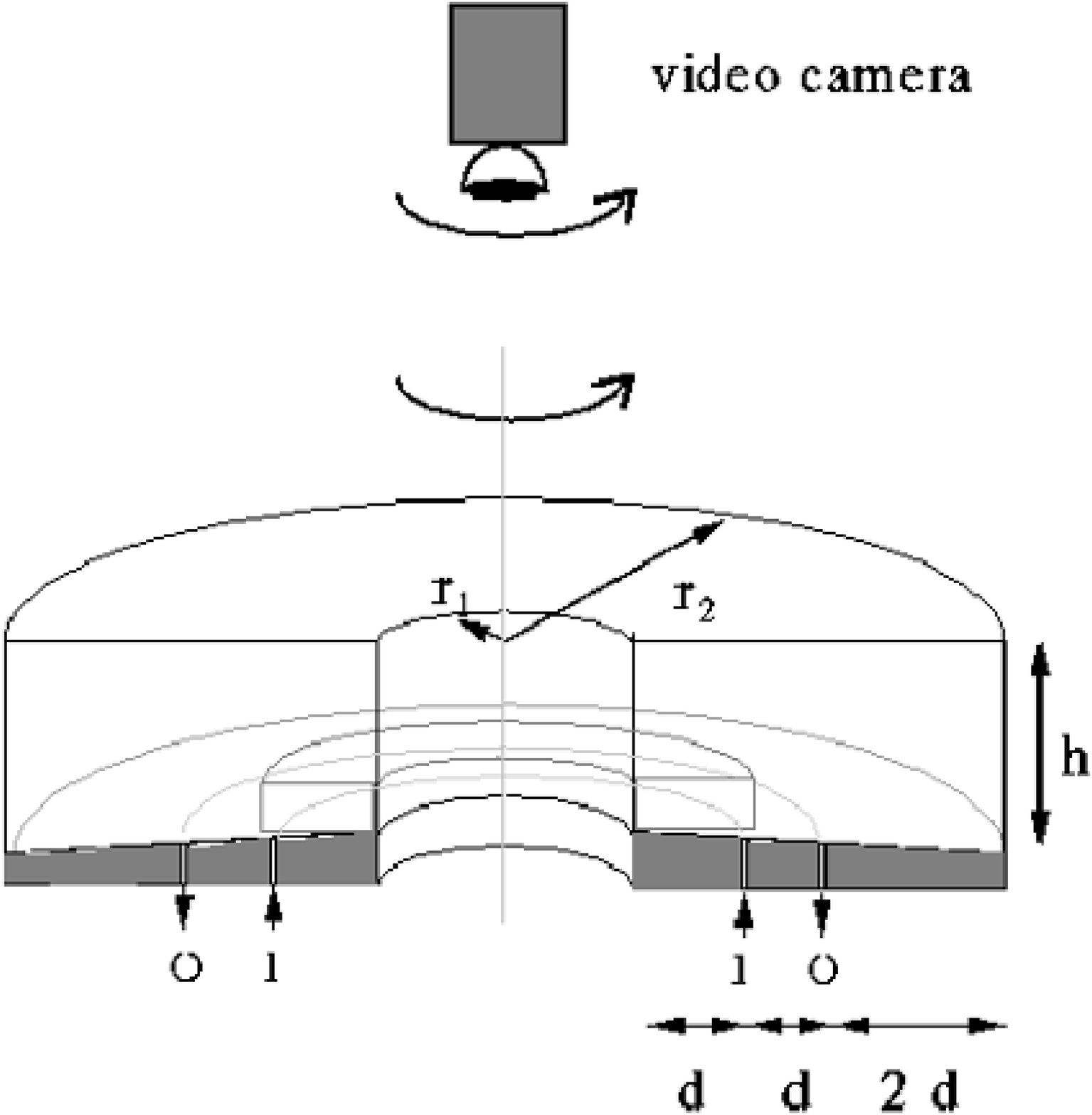}}\\\resizebox{0.35\columnwidth}{!}{
  \includegraphics{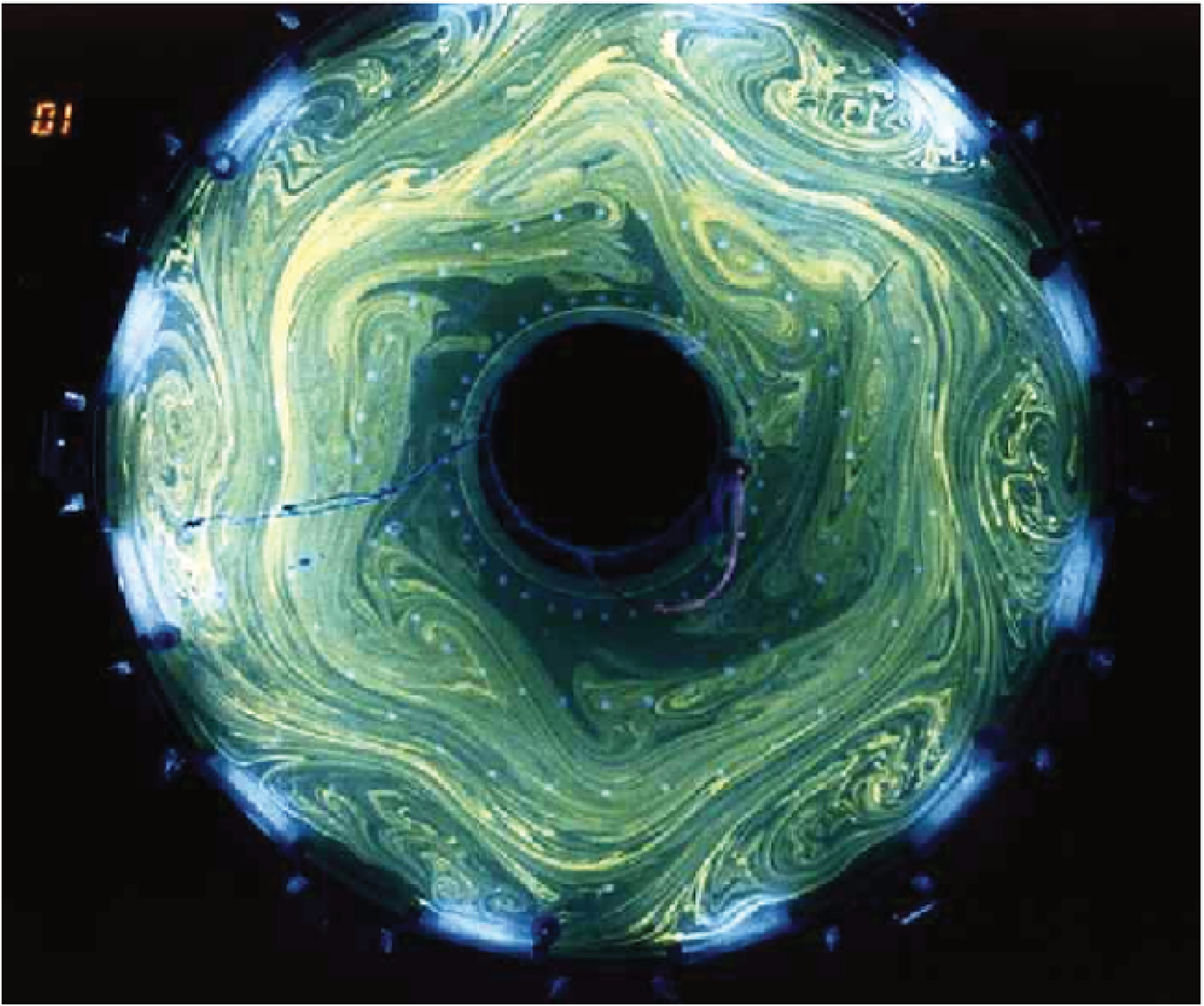}}
  \resizebox{0.35\columnwidth}{!}{  \includegraphics{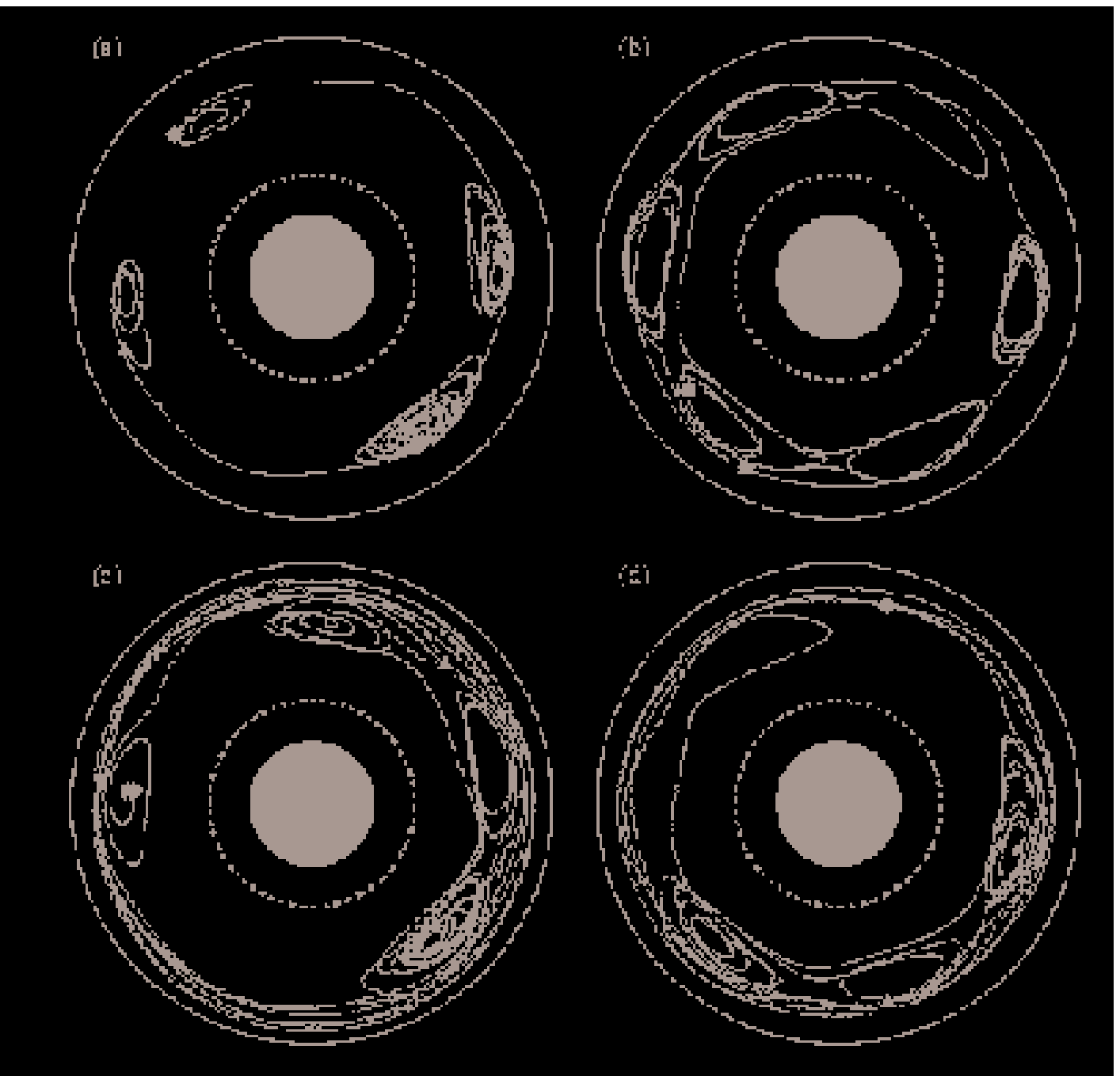}}
  \caption{(a) Rotating annulus(b) The formation of eddies inside the rotating annulus, as
recorded by the camera (left panel), and (b) typical orbits of tracer particles
inside the annulus (right panel).}\label{anorbits}
\end{figure}
The simple experiment of the rotating annulus, shown in Fig. \ref{anorbits}a,
allows to illustrate
the differences between normal and
anomalous diffusion \cite{Solomon94,Weeks96}.
Water is pumped into the annulus through a ring of holes marked with $I$
and pumped out through a second ring of holes marked with $O$.
The annulus is completely
filled with water and rotates as a rigid body (the inner and outer walls
rotate together). The pumping of the fluid generates a turbulent
flow in the annulus.
A camera on top of the annulus
records the formation of the turbulent eddies inside the rotating annulus and
allows to track seeds of
different tracer particles injected into the fluid and to
monitor their orbits
(see Fig. \ref{anorbits}).

In the case of normal diffusion, which
occurs mainly in fluids close to equilibrium,
the particle trajectories are characterized by irregular,
but small steps, which makes trajectories look
irregular but still homogeneous.
The trajectories shown
in Fig.\ \ref{anorbits} for the highly turbulent rotating annulus,
which is far away from equilibrium,
show different types of orbits, with two basic new characteristic,
there is ``trapping" of particles
inside the eddies, where particles stay for 'unusually' long times
in a relatively small spatial area,
and there are ``long flights" of particles, where particles
are carried in a short time
step over large distances, in some cases almost through th entire system (for details on normal and anomalous diffusion of particles  see \cite{Isliker08}).

\subsection{Charged particle diffusion in a fractal distribution of UCS}

It has been suggested already that the coronal part of the complex magnetic topology above the AR is densely populated by UCS. Their spatial population is fractal \cite{Dim09} and their overall statistical characteristics e.g Probability Distribution Functions (PDF) of Volumes ($P(V))$ and Currents ($P(J)$)  can be estimated from the extrapolated magnetic topologies or the SOC models when they reach the SOC state \cite{Isl00,Dim13,tout}.

Vlahos et al. \cite{Vlahos04} performed a Monte Carlo simulation of the extended Continuous Time Random Walk \cite{Isliker08},
in position and momentum space, in application to flares in the solar
corona, with particular interest in the heating and acceleration of ions and electrons.
\begin{figure}[htb]
\centering
\resizebox{0.45\columnwidth}{!}{  \includegraphics{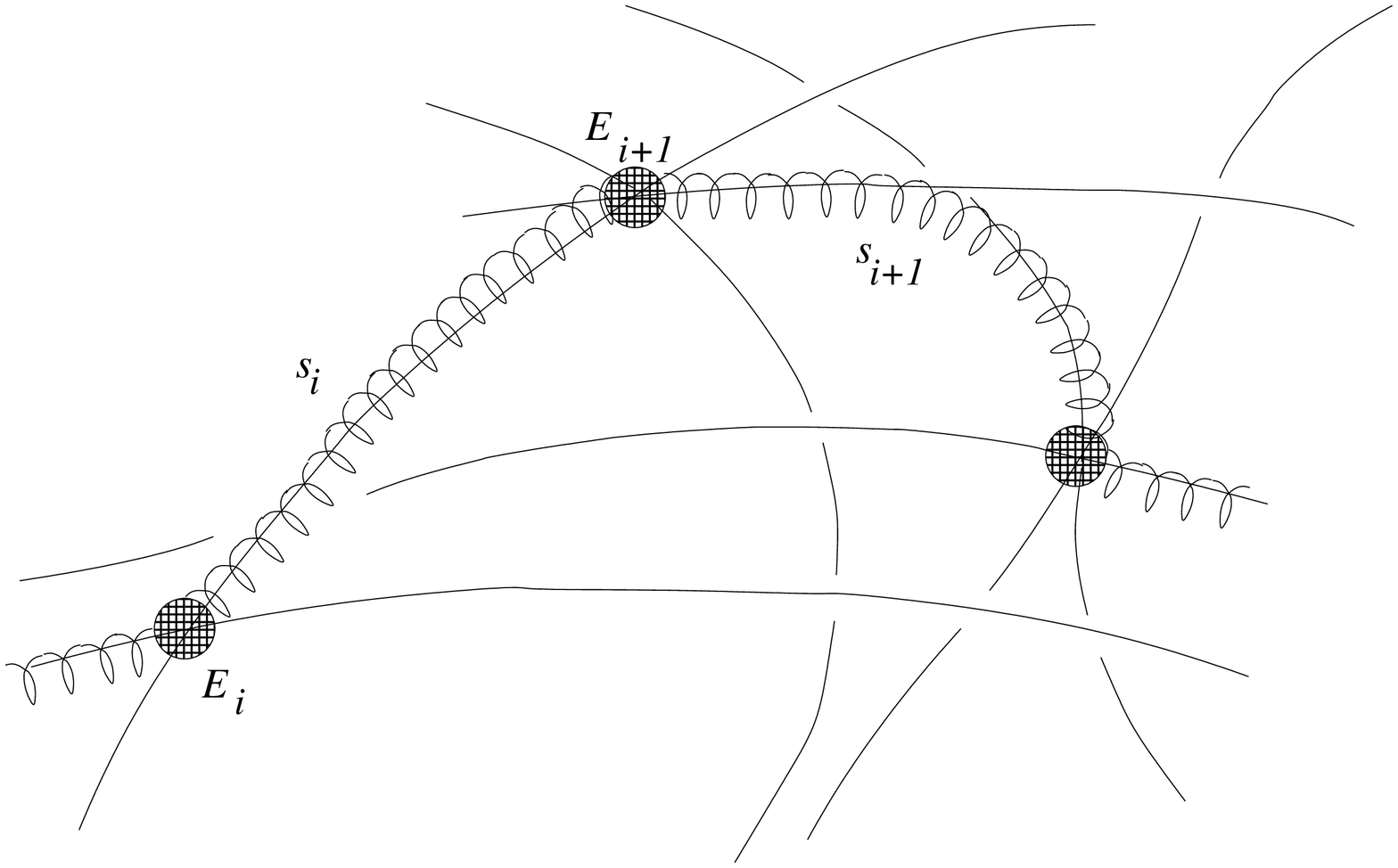}}
\resizebox{0.45\columnwidth}{!}{\includegraphics{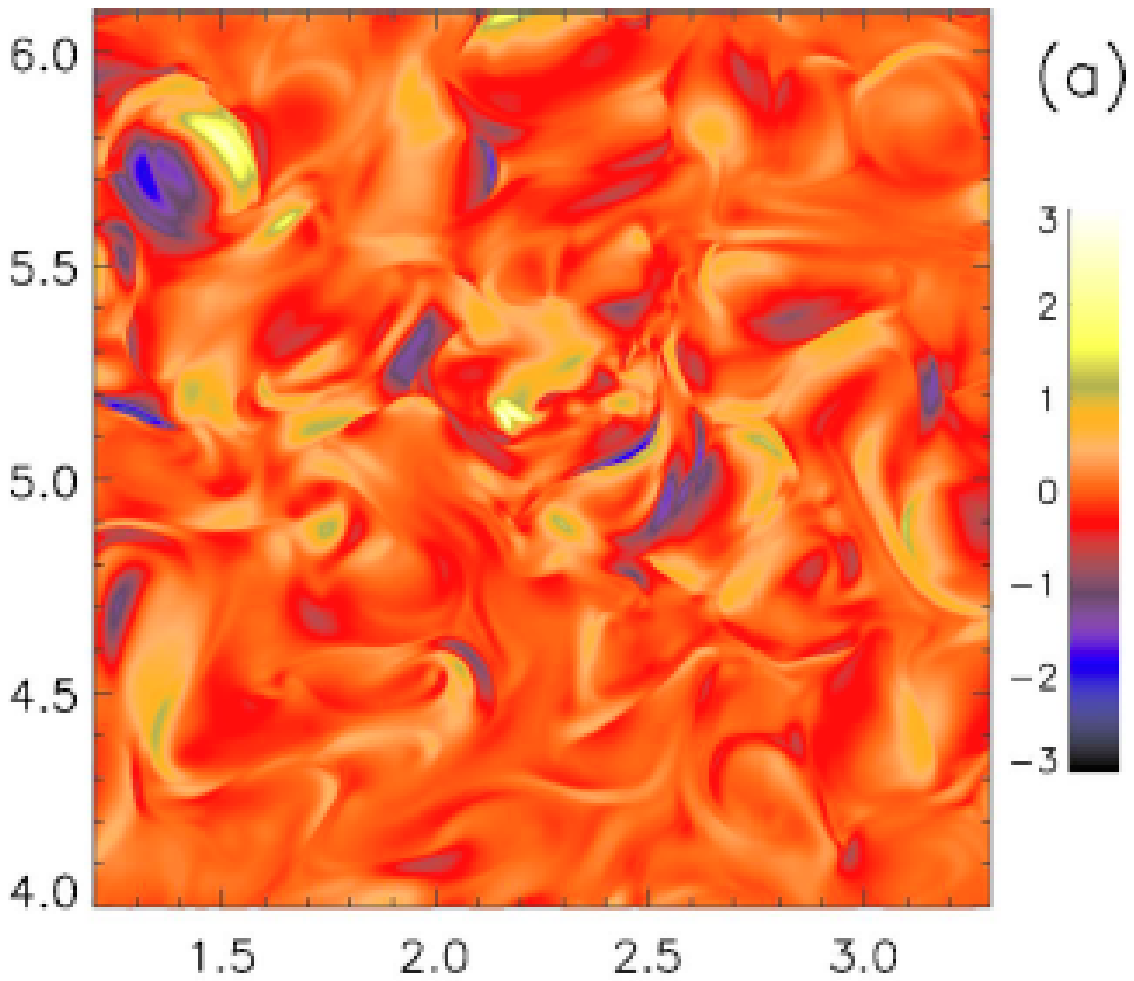}}
  \caption{(a) The basic elements of the random walk in a fractal collection of UCS \cite{Vlahos04},  (b) Contour plots of the total electric field contribution from 2D MHD turbulence \cite{Servi10}}\label{atr}
\end{figure}

The motion of charged particles inside an environment of randomly distributed UCS (see Fig. \ref{atr})  can be analyzed with the use of the two PDFs $P(V), P(J)$  and the fractal dimension $D_F$, following methods developed by Vlahos et al. \cite{Vlahos04}. The charged particle (electron or ion) starts at a random point inside the AR with a random velocity along the magnetic field $u_i$. The initial velocity distribution of the particles is a Maxwellian with initial temperature $T$. The ambient density of the  particles in the low corona is approximately constant $n_0$. The charged particle moves freely along a distance $s_i$ estimated from the fractal dimension $D_F$  (see details below)
until it reaches a current sheet where it is energized  by the electric field. This is estimated by Ohm's law $E=\eta J$, where $\eta$ is the local resistivity and $J$ is estimated from the probability distribution $P(J)$. In the case that the free travel of particles is longer than the collisional mean free path  we include the collisional losses in our analysis. We follow the evolution of the particle distribution  in successive time intervals.  Let us discuss briefly below the way we reconstruct the dynamic evolution of a distribution of particles inside a fractal distribution of current sheets.

\paragraph{Free travel distance.$-$}\label{ftt}
As it was pointed out by \cite{Isl03}, the probability of a particle, starting at an UCS in the AR, to travel freely a distance $s$ before meeting again an UCS is
\begin{equation}\label{frac}
P(s)=\frac{D_F-2}{s_{\rm max}^{D_F-2}-s_{\rm min}^{D_F-2}}s^{D_F-3}\;\;\;\; s_{\rm min}<s<s_{\rm max}
\end{equation}
if the UCSs are fractally distributed.  This formula is an approximation that applies if $ D_F$ is strictly smaller than 2, as for the case examined here, where the distribution is of power-law shape (the corresponding expressions for the cases $D_F=2$ and $D_F>2$ are different and not cited here).  Using the distribution $P(s)$, we generate sequences of random free travel distances $s_i$.

\paragraph{Collisional losses.$-$}\label{parcollisions}
The electron and ion Coulomb collision frequency 
is given by
\begin{equation}\label{colfr}
\nu_e = \frac{4\pi n e^4 \ln\Lambda }{m_j^2 V_j^3}
\end{equation}
where $e$ is the elementary charge, $m_e$ the electron mass, $m_i$ the ion mass and $\Lambda$ the Coulomb logarithm  (see \cite{Karney}) $n$ the number density, $j=e,i$ and $V_j$ the thermal velocity.  The particles thus loose energy as they travel between current sheets located at distances $s_i,$ that are much larger than the mean free path.

\paragraph{Acceleration length.$-$}\label{al}
Assuming that the particles interact with a current sheet with volume $V_j=\ell_j^2 \times d$, where $d$ is the width of the UCS , we estimate easily the length $\ell_j$ of the current sheet,
\begin{equation}\label{acc}
\ell_j=\sqrt{\frac{V_j}{d}}
\end{equation}
where the volume $V_j$ follows the probability distribution
\begin{equation}\label{vol}
P(V_j)=AV_j^{-a}, \;\;\; V_j^{min}< V_j< V_j^{max},
\end{equation}
where $A$ is the normalization constant,
Combining Eqs. \ref{acc} and \ref{vol} we can estimate the UCSs'
random length $\ell_j$.

\paragraph{Electric field strength.$-$}\label{ef}
The electric field along the magnetic field, as we mentioned already, inside  the current sheet is
\begin{equation}\label{Ohm}
E=\eta J
\end{equation}
where $J$ is the current given by the probability distribution $P(J)$. The resistivity is assumed close to zero when $J<j_{\rm th}$ and $\eta \approx \bar{\eta} \;\eta_S$ when $J>j_{\rm th}$, where $\eta_S $ is the Spitzer resistivity
\begin{equation}\label{spitzer}
\eta_S=\frac{m_e \nu_e}{n e^2}
\end{equation}
and $\bar{\eta}$ is a free parameter.
By including $\bar{\eta}$, we implicitly assume that due to the relatively strong currents ($J>j_{\rm th}$) low frequency electrostatic waves are excited and the particles interact with the waves much more efficiently than via Coulomb collisions, so the resistivity is enhanced by several orders of magnitude and is called `anomalous' \cite{Sagdeev,Papadop77,Ugai,Petkaki}. According to the literature stated above, $\bar{\eta}$ is proportional to $(J-j_{\rm th}).$  The choices for $\bar{\eta}$ can be very different
for flux emergence and explosive events, thus affecting dramatically both the time
scale of energy release and the heating and acceleration of particles.

\paragraph{Equations of Motion.$-$}\label{trajectory}
We assume that the motion is one dimensional along the magnetic field lines and the velocity of the particles generally is relativistic. The motion of the particle is divided into two parts:

 (a) Free travel
along a distance $s_i$, suffering only collisional losses,
where we apply the simplified model of \cite{Lenard58} for
the Coulomb collisions of charged particles with a background plasma population
of temperature $T_b$,
\begin{equation}\label{loss1}
  \frac{ds}{dt}=u
\end{equation}
\begin{equation}\label{loss}
 \frac{du}{dt}=- \nu_{e} u + \left (\sqrt{2\nu_{e}k_B T_b/m_j} \right )\, W_t
\end{equation}
where  $k_B$ is the Boltzmann constant, $m_j$ the particle (ion/electron) mass, and $W_t$
is an independent Gaussian random variable with
mean value zero and variance the integration time-step $\Delta t$.
Equations (\ref{loss1}) and (\ref{loss}) are solved by directly using the analytical solution
$s_a(\tau;s_0,u_0)$  and $u_a(\tau;u_0)$ (with $s_0$ and $u_0$ the values of the position and velocity at $\tau=0$) (see \cite{Gillespie96}). For a prescribed free travel distance $s_i$, we first calculate the total free travel time $\tau_i$ by solving the nonlinear equation $s_a(\tau_i; s_0=0,u_0=u(t))=s_i$, and then determine the new velocity as $u(t+\tau_i)=u_a(\tau_i;u_0=u(t))$ in one step. This method has the additional benefit that it
allows to make the collision model
to be more realistic in that the collision frequency can be made
proportional to $1/u^3$, with the characteristic reduced collisionality
at high velocities.

(b) The particle is energized  by the UCS of length $\ell_j$ (Eq. \ref{acc})
\begin{equation}\label{space2}
  \frac{ds}{dt}=u
\end{equation}
\begin{equation}\label{cs}
\frac{du}{dt}=(e/m_j) \cos(\alpha) E
\end{equation}
with $\alpha$ a random angle between the magnetic and the
electric field, in order to take into account the arbitrary directionality of the electric field.

The particle motion inside a turbulent reconnecting volume, with a large number of UCS    present (Fig. \ref{atr}) has many similarities with the first order Fermi   acceleration and it is a very efficient accelerator (see \cite{Dahlin}).

\section{Summary and Discussion}
\label{sec:5}

The last twenty five years several  attempts have been made to explore the links between the convection zone and the AR as a multiscale turbulent laboratory. These attempts started some time ago, when the 3D MHD simulations were in their infancy (see for example Fig. 1 in \cite{Anastasi94} and Fig. 1 in \cite{Vlahos04}). We have returned to this theme in this review with a new and improved synthesis which now relies more on MHD simulations and less on CA models as it was the case several years ago. We are analysing in this review a turbulent multi-scale system:

\begin{enumerate}
	\item The global scale under study in this review is the solar AR and it has a scale of thousands of Mm (see Fig. \ref{AR5}b). This global interaction of the convection zone with the AR is responsible for the formation of the thin magnetic flux tubes and their transport to the solar surface where the photosheric part of the AR is formed.
	\item Using the NLFF techniques for the reconstruction of the complex topology above the AR we explore the formation of the UCS (see Fig. \ref{vl1}b). With the help of the 3D MHD codes we can explore the formation and evolution of the UCS inside a box with a scale of several Mm \cite{Dmitruk03}.
	\item Dropping the scale to thousands of Km, we can simulate the  collection of UCS which are in close proximity (see Fig. \ref{CS1}b) and can {\bf interact} nonlinearly \cite{Gal96}	
	\item Moving down to tens of meters we observe the evolution and fragmentation of the isolated UCS (see Fig. \ref{sfrag}a)  \cite{Dahlin}.  	
	\end{enumerate}

We thus here perform the analysis of the complex topology of the 3D magnetic field above the AR using methods borrowed from the complexity theory and 3D MHD and/or Kinetc simulations. The main steps followed in our review are:

\begin{enumerate}
	\item {\bf Formation of AR}: On the largest scales (Fig. \ref{AR5}b)  the convection zone played the main role in forming (dynamo) and injecting and randomly perturbing {\bf the magnetic flux tubes} once they have been emerged above the photosphere. The use of the percolation theory in conjunction with the 3D MHD is an obvious step to explore the statistical characteristics of this interaction but the progress in the cross talking of the 3D MHD models with models based on the percolation theory  has been very slow. In the language of SOC this step is called {\bf the driver}.
	\item {\bf Formation of magnetic discontinuities and UCS}: The reconstruction of the AR with the use of NLFF and the identification of UCS inside the 3D coronal part of the AR (Fig. \ref{vl1}b) has been explored by many numerical models. Parker in his famous conjecture \cite{Parker88} suggested that the random motion of the magnetic flux tubes will be responsible for the formation of the UCS. The formation of the UCS has also been explored by several 3D MHD simulations the last ten years \cite{Dmitruk03}. In the language of SOC this step is one of the rules of the CA model and is called {\bf Loading}.
	\item {\bf Multiple UCS as a host of avalanches} Reducing our resolution even further to tens of Km we can observe the presence of many UCS. The simulations of many interacting UCS and their evolution approaching asymptotically   {\bf avalanches} it have been analysed by Hood et al. \cite{Hood15} recently.
	\item {\bf Fragmentation of UCS and magnetic energy redistribution}: On the smallest scale (tens of meters) the fragmentation of a UCS has been established (see the 3D MHD and kinetic simulations in \cite{marco1,Dahlin}). Translating this step into the CA {\bf rules} means redistribution of magnetic energy and potentially drives avalanches on the larger scales, as we pointed out before.	
		\item {\bf Turbulent reconnection and particle acceleration} Once the system reaches the SOC state a collection of UCS will have  a fractal distribution   inside the AR. This state is analogous to the {\bf turbulently reconnecting state} reported in many turbulent systems in astrophysics \cite{Lazarian12} and has been explored with several 3D numerical codes.  This stage is strongly coupled with the current fragmentation and the appearance   of avalanches in the evolution of the UCS. In the language of SOC this is called {\bf SOC state}. It has been proposed that the interaction of charged particles with the UCS \cite{Vlahos04} is an efficient accelerator. This approach to energy transport from the UCS to the plasma has many similarities and important differences with the well known Fermi type accelerators with the role of the {\bf magnetic clouds} being taken by the UCS. 		

\end{enumerate}

Reducing all  the above analysis into the rules of the CA model:
\begin{itemize}
	\item {\bf Driver} Convection zone turbulence and very large scales (thousands of Mm)
	\item {\bf Loading} Fractal photospheric patterns formed by the emergence and random motion of the magnetic flux tubes are driving the formation of UCS ( a few Mm scale)
	\item {\bf Local redistribution of energy} Fragmentation of UCS on  local scales of tens of meters
	\item {\bf Avalanches} from the interaction of UCS on all scale from several Kms to several Mms.
	\item {\bf The system reaches a SOC state asymptotically} The turbulent reconnection on the Mm scales.
	\item {\bf First order Fermi type heating and acceleration} The interaction of the particles with UCS in the turbulent reconnection stage.
	
\end{itemize}

It is important to stress in this review  that the connection of the {\bf turbulent driver} (convection zone) with the {\bf turbulent reconnection} state of the complex magnetic topology in the coronal part of the AR is not simple and the transfer of energy from the large scales to  the small scales {\b UCS} where it is dissipated  is not analogous to the simple Kolmogoroff ideas presented in the 40s. We are dealing with a different type  of coupled system and the transfer of energy from the large scales to the dissipation scales which may be much more common in large scale astrophysical systems.

 Unfortunately the research groups  working with the MHD and kinetic codes have not established yet a familiarity with the methods of complexity presented in this review and the transfer of knowledge between the two communities  is relatively  slow. We belive that this gap will close in the coming years.



%
%
\vspace{0.5cm}
{\bf Acknowledgements:} We thank Drs A. Anastasiadis and M. Georgoulis for reading the article and making valuable suggestions.  authors acknowledge support by the European Union (European Social Fund -ESF) and Greek national funds through the Operational Program Education and Lifelong Learning of the National Strategic Reference Framework (NSRF) -Research Funding Program: Thales: Investing in knowledge society through the European Social Fund.

%
%

\end{document}